\documentclass[12pt]{article}
\usepackage{epsfig}

\def\beq{\begin{equation}}
\def\eeq#1{\label{#1}\end{equation}}
\def\eeqn{\end{equation}}

\def\beqa{\begin{eqnarray}}
\def\eeqa#1{\label{#1}\end{eqnarray}}
\def\eeqan{\end{eqnarray}}

\let\bar=\overbar

\def\Dslash{\not{\hbox{\kern-4pt $D$}}}
\def\dslash{\not{\hbox{\kern-2pt $\del$}}}

\def\msb{{\bar{\ssstyle M \kern -1pt S}}}

\def\Title#1{\begin{center} {\Large {\bf #1} } \end{center}}

\begin{document}

%------------------------BEGINNING OF CERN TITLE PAGE----------------------

\thispagestyle{empty}

\begin{flushright}
CERN-PH-TH/2007-053\\
hep-ph/0703112
\end{flushright}

\vspace{2.0truecm}
\begin{center}
\boldmath
\large\bf CP Violation and B Physics at the LHC
\unboldmath
\end{center}

\vspace{0.9truecm}
\begin{center}
Robert Fleischer\\[0.1cm]
{\sl Theory Division, Department of Physics, CERN\\
CH-1211 Geneva 23, Switzerland}
\end{center}

\vspace{0.9truecm}

\begin{center}
{\bf Abstract}
\end{center}

{\small
\vspace{0.2cm}\noindent
In this decade, there are huge efforts to explore $B$-meson decays, which provide 
an interesting playground for stringent tests of the Standard-Model description of the
quark-flavour sector and the CP violation residing there. Thanks to the $e^+e^-$ $B$ 
factories at KEK and SLAC, CP violation is now a well-established phenomenon in the 
$B$-meson system, and recently, also $B^0_s$--$\bar B^0_s$ mixing could be 
measured at the Tevatron. The decays of $B^0_s$ mesons are the key target of the 
$B$-physics programme at the LHC, and will be the focus of this presentation, 
discussing the theoretical aspects of various benchmark channels and the question 
of how much space for new-physics effects in their observables is left by the recent 
experimental results.
}

\vspace{0.9truecm}

\begin{center}
{\sl Invited talk at the 8th International Workshop on Heavy Quarks and Leptons\\
Deutsches Museum, Munich, Germany, 16--20 October 2006\\
To appear in the Proceedings}  
\end{center}

\vfill
\noindent
CERN-PH-TH/2007-053\\
March 2007

\newpage
\thispagestyle{empty}
\vbox{}
\newpage
 
\setcounter{page}{1}

%------------------------END OF CERN TITLE PAGE------------------------------

\Title{CP Violation and B Physics at the LHC}

%\begin{center}{\large \bf Contribution to the proceedings of HQL06,\\
%Munich, October 16th-20th 2006}\end{center}

\bigskip\bigskip

%+\addtocontents{toc}{{\it D. Reggiano}}
%+\label{ReggianoStart}

\begin{raggedright}  

{\it Robert Fleischer\index{Fleischer, R.}\\
Theory Division, Department of Physics, CERN\\
CH-1211 Geneva 23, SWITZERLAND}
\bigskip\bigskip
\end{raggedright}

\section*{Abstract}
In this decade, there are huge efforts to explore $B$-meson decays, which provide 
an interesting playground for stringent tests of the Standard-Model description of the
quark-flavour sector and the CP violation residing there. Thanks to the $e^+e^-$ $B$ 
factories at KEK and SLAC, CP violation is now a well-established phenomenon in the 
$B$-meson system, and recently, also $B^0_s$--$\bar B^0_s$ mixing could be 
measured at the Tevatron. The decays of $B^0_s$ mesons are the key target of the 
$B$-physics programme at the LHC, and will be the focus of this presentation, 
discussing the theoretical aspects of various benchmark channels and the question 
of how much space for new-physics effects in their observables is left by the recent 
experimental results.

\section{Setting the Stage}
In the Standard Model (SM), the phenomenon of CP violation can be accommodated 
in an efficient way through a complex phase entering the quark-mixing matrix,
which governs the strength of the charged-current interactions of the quarks 
\cite{KM}. This Kobayashi--Maskawa (KM) mechanism of CP violation is the 
subject of detailed investigations in this decade. The main interest in the study
of CP violation and flavour physics in general is due to the fact that new physics 
(NP) typically leads to new patterns in the flavour sector. This is actually the case in 
several specific extensions of the SM, such as SUSY scenarios, left--right-symmetric 
models, models with extra $Z'$ bosons, scenarios with extra dimensions, or 
``little Higgs" scenarios. Moreover, also the observed neutrino masses point towards an
origin lying beyond the SM, raising now also the question of having CP violation 
in the neutrino sector and its connection with the quark-flavour physics. Finally, the
baryon asymmetry of the Universe also suggests new sources of CP violation. These
could be associated with very high energy scales, where a particularly interesting
scenario is provided by ``leptogenesis" \cite{BPY}, involving typically new 
CP-violating sources in the decays of heavy Majorana neutrinos. On the other hand, 
new CP-violating effects arising in the NP scenarios listed above could in fact be 
accessible in the laboratory. 

Before searching for signals of NP, we have first to understand the SM picture.
Here the key problem is due to the impact of strong interactions, leading to
``hadronic" uncertainties. A famous example is the quantitiy 
$\mbox{Re}(\varepsilon'/\varepsilon)_K$, which measures the direct CP violation
in neutral $K$ decays (for an overview, see \cite{buja}). In the kaon system,
where CP violation was discovered in 1964 \cite{CPV}, clean tests of the SM are 
offered by the decays $K^+\to\pi^+\nu\bar\nu$ and $K_{\rm L}\to\pi^0\nu\bar\nu$, 
where the
hadronic pieces can be fixed through $K\to\pi\ell\bar\nu$ modes. These rare 
decays are absent at the tree level of the SM, i.e.\ originate there exclusively from 
loop processes, with resulting tiny branching ratios at the $10^{-10}$ level (for a 
recent review, see \cite{BSU}). Experimental studies of these channels are 
therefore very challenging. Nevertheless, there are plans to measure 
$K^+\to\pi^+\nu\bar\nu$ at the SPS (CERN) \cite{Ruggiero}, and efforts to explore 
$K_{\rm L}\to\pi^0\nu\bar\nu$ at the E391 (KEK/J-PARC) experiment. 

The $B$-meson system is a particularly promising probe for the testing of the
quark-flavour sector of the SM, and will be the focus of this presentation. It offers
various strategies, i.e.\ simply speaking, there are many $B$ decays that we can 
exploit, and we may search for clean SM relations that could be spoiled through
the impact of NP. There are two kinds of experimental facilities, where $B$-meson
decays can be studied:
\begin{itemize}
\item The ``$B$ factories", which are asymmetric $e^+e^-$ colliders operated at
the $\Upsilon(4S)$ resonance, producing only $B^0_d\bar B^0_d$ and
$B^+_uB^-_u$ pairs: PEP-II with the Babar experiment (SLAC) and KEK-B with the 
Belle experiment (KEK) have by now produced altogether ${\cal O}(10^9)$ $B\bar B$ pairs, establishing CP violation in the $B$ system and leading to many other 
interesting results. There are currently discussions of a super-$B$ factory, with an 
increase of luminosity by two orders of magnitude~\cite{Iijima}. 
\item Hadron colliders produce, in addition to $B_d$ and $B_u$,  also $B_s$ 
mesons,\footnote{Recently, data were taken by Belle at $\Upsilon(5S)$, allowing also access to $B_s$ decays \cite{Belle-U5S}.} as well as $B_c$ and $\Lambda_b$ 
transitions. The Tevatron experiments CDF and D0 have reported first 
$B_{(s)}$-decay results. The physics potential of the $B_s$ system can be fully 
exploited at the LHC, starting operation in autumn 2007. Here ATLAS and CMS can 
also address some $B$ physics topics, although these studies are the main target 
of the dedicated $B$-decay experiment LHCb.
\end{itemize}
The central target of these explorations is the well-known unitarity triangle (UT) of the
Cabibbo--Kobyashi--Maskawa (CKM) matrix with its three angles $\alpha$, $\beta$
and $\gamma$. Its apex is given by the generalized
Wolfenstein parameters \cite{wolf-blo}
\begin{equation}%\label{}
\bar\rho \equiv (1-\lambda^2/2)\rho, \quad
\bar\eta \equiv (1-\lambda^2/2)\eta.
\end{equation}

The key processes for the exploration of CP violation are given by non-leptonic decays 
of $B$ mesons, where only quarks are present in the final states. In these transitions,
CP-violating asymmetries can be generated through interference effects. Depending
on the flavour content of their final states, non-leptonic $B$ decays receive 
contributions from tree and penguin topologies, where we distinguish between
QCD and electroweak (EW) penguins in the latter case. The calculation of the
decay amplitudes, which can be written by means of the operator product 
expansion as follows \cite{BBL}:
\begin{equation}%\label{}
A(B\to f)\sim \sum\limits_k
\underbrace{C_{k}(\mu)}_{\mbox{pert.\ QCD}} 
\times\,\,\, \underbrace{\langle f|Q_{k}(\mu)|B\rangle}_{\mbox{``unknown''}},
\end{equation}
remains a theoretical challenge, despite interesting recent progress \cite{beneke}.

\begin{figure}[t]
\begin{center}
\begin{tabular}{ll}
    \includegraphics[width=6.8truecm]{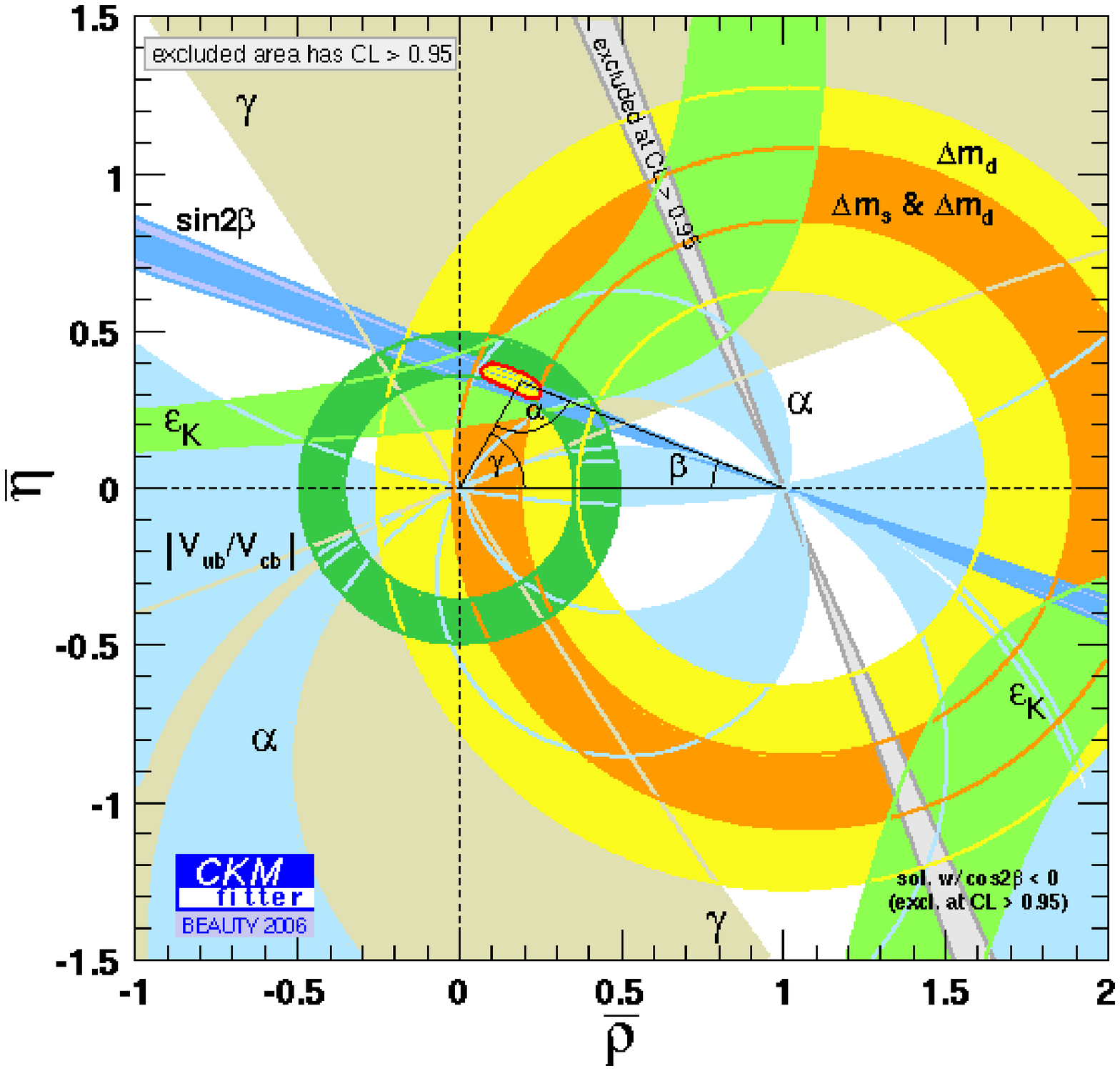}  &
    \hspace*{0.0truecm} \includegraphics[width=7.3truecm]{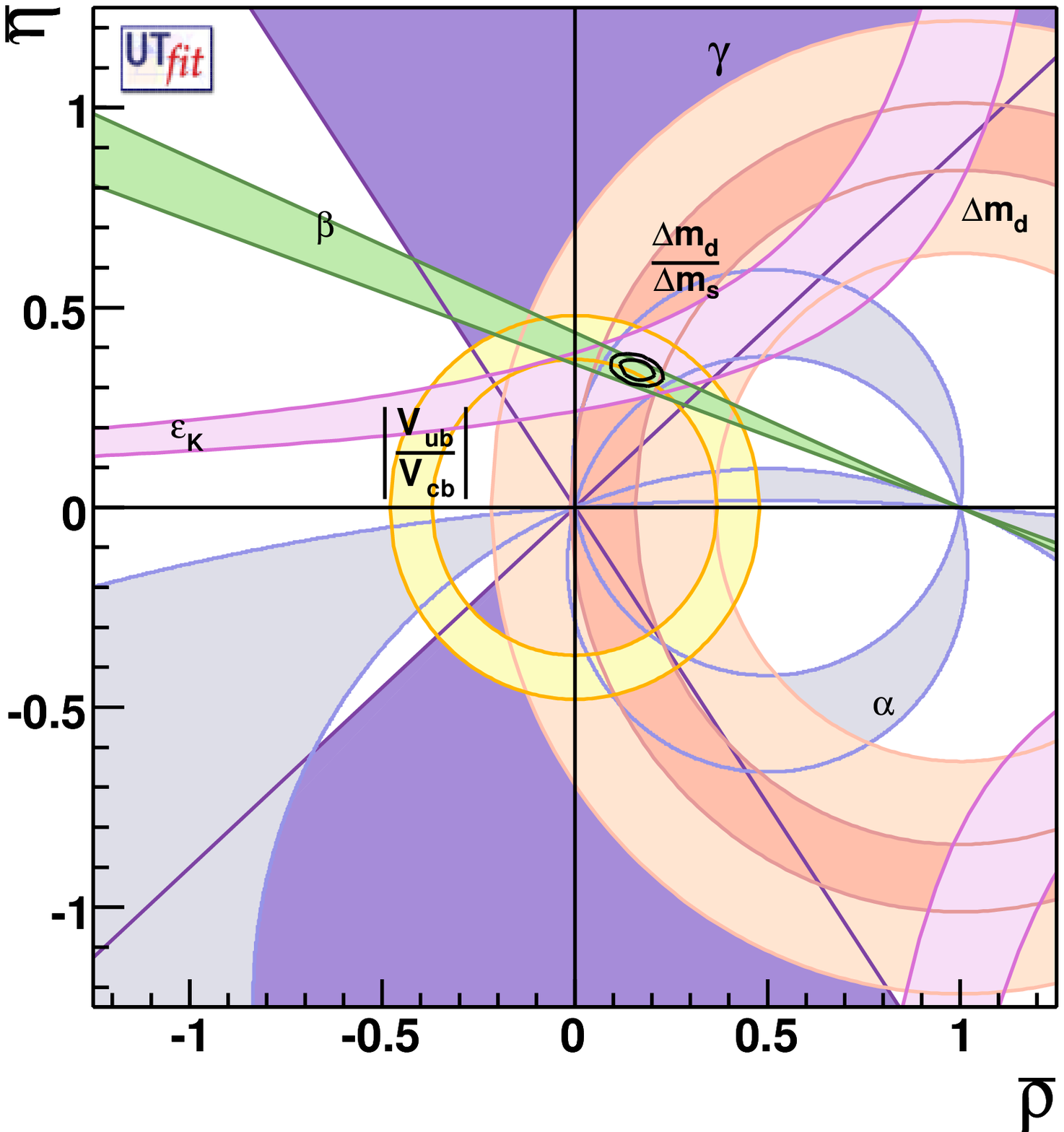} 
   \end{tabular}
   \end{center}
   \vspace*{-0.6truecm}
\caption{Analyses of the CKMfitter and UTfit collaborations \cite{CKMfitter,UTfit}.}
\label{fig:UT-status}
\end{figure}

However, for the exploration of CP violation, the calculation of the hadronic matrix 
elements $\langle f|Q_{k}(\mu)|B\rangle$ of local four-quark operators can actually 
be circumvented. This feature is crucial for a stringent testing of the CP-violating
flavour sector of the SM. To this end, we may follow two avenues:
\begin{itemize}
\item Amplitude relations allow us in fortunate cases to eliminate the hadronic 
matrix elements. Here we distinguish between exact relations, 
using pure ``tree'' decays  of the kind $B^\pm\to K^\pm D$ \cite{gw,ADS} or 
$B_c^\pm\to D^\pm_s D$ \cite{fw}, and relations, which follow from the flavour symmetries of strong interactions, i.e.\ isospin or $SU(3)_{\rm F}$, and 
typically involve $B_{(s)}\to\pi\pi,\pi K,KK$ modes~\cite{GHLR}. 
\item In decays of neutral $B_q$ mesons ($q\in\{d,s\}$), the interference between
$B^0_q$--$\bar B^0_q$ mixing and $B^0_q, \bar B^0_q\to f$ decay processes 
leads to ``mixing-induced" CP violation. If one CKM amplitude dominates the
decay, the essentially ``unknown" hadronic matrix elements cancel. The key 
application of this important feature is the measurement of $\sin2\beta$ through the 
``golden" decay $B^0_d\to J/\psi K_{\rm S}$ \cite{bisa}.  
\end{itemize}

Following these lines, various processes and strategies emerge for the exploration
of CP violation in the $B$-meson system (for a more detailed discussion, see
\cite{RF-lect}). In particular, decays with a very different dynamics allow us to
probe the same quantities of the UT. These studies are complemented by rare 
decays of $B$ and $K$ mesons \cite{weiler}, which originate from loop processes in 
the SM model and show interesting correlations with the CP violation in the $B$ system. 
In the presence of NP, discrepancies should show up in the resulting roadmap of 
quark-flavour physics. 

In Fig.~\ref{fig:UT-status}, we show the current status of the UT \cite{stocchi}
emerging from the comprehensive -- and continuously updated -- analyses by 
the ``CKM Fitter Group'' 
\cite{CKMfitter} and the ``UTfit collaboration'' \cite{UTfit}. We observe that there
is impressive global agreement with the KM mechanism. However, there is also 
some tension present, as the straight line representing the measurement of 
$(\sin 2\beta)_{\psi K_{\rm S}}$ is now on the lower side of the UT side $R_b$
measured through $|V_{ub}/V_{cb}|$. We shall return to this topic in
Section~\ref{ssec:NP-Bd}. Let us next discuss the interpretation of the $B$-factory 
data in more detail.

\boldmath
\section{A Brief Look at the Current $B$-Factory Data}
\unboldmath
There are two popular avenues for NP to manifest itself in the $B$-factory data:
through effects entering at the decay amplitude level, or through $B^0_q$--$\bar B^0_q$
mixing. 

\subsection{New Physics at the Decay Amplitude Level}
If a given decay is dominated by SM tree processes, we have typically small effects
through NP contributions to its transition amplitude. On the other hand, we may 
have potentially large NP effects in the penguin sector through new particles in 
the loops or new contributions at the tree level (this may happen, for instance, 
in SUSY or models with extra $Z'$ bosons). The search for such signals of NP 
in the $B$-factory data has been a hot topic for several years. 

\boldmath
\subsubsection{CP Violation in $b\to s$ Penguin Modes}\label{ssec:bsss}
\unboldmath
A particularly interesting probe of NP is the decay $B^0_d\to \phi K_{\rm S}$. It is
caused by $\bar b\to \bar s s \bar s$ quark-level processes, i.e.\ receives only 
contributions from penguin topologies. The corresponding final state is 
CP-odd, and the time-dependent CP asymmetry takes the following
form:\footnote{We shall use a similar sign convention also for 
self-tagging neutral $B_d$ and charged $B$ decays.}
\begin{eqnarray}%\label{}
\lefteqn{\frac{\Gamma(B^0_d(t)\to \phi K_{\rm S})-
\Gamma(\bar B^0_d(t)\to \phi K_{\rm S})}{\Gamma(B^0_d(t)\to \phi K_{\rm S})+
\Gamma(\bar B^0_d(t)\to \phi K_{\rm S})}}\nonumber\\
&&={\cal A}_{\rm CP}^{\rm dir}(B_d\to\phi K_{\rm S})\,\cos(\Delta M_d t)+
{\cal A}_{\rm CP}^{\rm mix}(B_d\to\phi K_{\rm S})\,\sin(\Delta M_d t),\label{ACP-timedep}
\end{eqnarray}
where ${\cal A}_{\rm CP}^{\rm dir}(B_d\to\phi K_{\rm S})$ and 
${\cal A}_{\rm CP}^{\rm mix}(B_d\to\phi K_{\rm S})$ denote the direct and
mixing-induced CP asymmetries, respectively. Thanks to the weak phase structure
of the $B^0_d\to \phi K_{\rm S}$ decay amplitude in the SM, we obtain the 
following expressions \cite{RF-lect}: 
\begin{eqnarray}
{\cal A}_{\rm CP}^{\rm dir}(B_d\to \phi K_{\rm S})&=&0+
{\cal O}(\lambda^2)\label{BphiK-rel1}\\
{\cal A}_{\rm CP}^{\rm mix}(B_d\to \phi K_{\rm S})&=&-\sin\phi_d
+{\cal O}(\lambda^2),\label{BphiK-rel2}
\end{eqnarray}
where $\phi_d$ is the $B^0_d$--$\bar B^0_d$ mixing phase and the 
doubly Cabibbo-suppressed ${\cal O}(\lambda^2)$ terms describe 
hadronic corrections. Since the mixing-induced CP asymmetry of the
$B_d\to J/\psi K_{\rm S}$ channel measures also $-\sin\phi_d$, we arrive at 
the following SM relation \cite{RF-EWP-rev,growo}:
\begin{equation}\label{Bd-phiKS-SM-rel}
-(\sin2\beta)_{\phi K_{\rm S}}\equiv
{\cal A}_{\rm CP}^{\rm mix}(B_d\to \phi K_{\rm S}) 
={\cal A}_{\rm CP}^{\rm mix}(B_d\to J/\psi K_{\rm S}) + 
{\cal O}(\lambda^2),
\end{equation}
which offers an interesting test of the SM. Since $B_d\to \phi K_{\rm S}$ is 
dominated, in the SM, by QCD penguin processes and receives significant 
contributions from EW penguins as well, the relations in (\ref{BphiK-rel1}) and 
(\ref{Bd-phiKS-SM-rel}) may well be affected by NP effects. This follows
through field-theoretical estimates for generic NP in the TeV regime
\cite{FM-BphiK}, and is also the case for several specific extensions of
the SM (see, e.g., \cite{BphiK-NP}). Concerning the
current experimental status \cite{DuFe}, it can be summarized through the averages 
obtained by the ``Heavy Flavour Averaging Group" \cite{HFAG}:
\begin{equation}%\label{}
(\sin2\beta)_{\phi K_{\rm S}}=0.39\pm0.18, \quad
{\cal A}_{\rm CP}^{\rm dir}(B_d\to \phi K_{\rm S})=0.01\pm0.13.
\end{equation}
During the recent years, the Belle results for $(\sin2\beta)_{\phi K_{\rm S}}$ 
have moved quite a bit towards the SM reference value of 
\begin{equation}
-{\cal A}_{\rm CP}^{\rm mix}(B_d\to J/\psi K_{\rm S}) \equiv
(\sin2\beta)_{\psi K_{\rm S}}=0.674\pm0.026, 
\end{equation}
and are now, within the errors, in agreement with the BaBar findings. Interestingly, 
the mixing-induced CP asymmetries of other $b\to s$ penguin modes show the 
same trend of having central values that are smaller than 0.674 \cite{HFAG}. This
feature may in fact be due to the presence of NP contributions to the corresponding
decay amplitudes. However, the large uncertainties do not yet allow us to draw
definite conclusions.

\boldmath
\subsubsection{The $B\to\pi K$ Puzzle}\label{ssec:BpiK}
\unboldmath
Another hot topic is the exploration of $B\to\pi K$ decays. Thanks to the $B$ factories, we could obtain valuable insights into these decays, raising the possibility of having 
a  modified EW penguin sector through the impact of NP, which has received a lot 
of attention in the literature (see, e.g., \cite{BpiK-papers}). Here we shall
discuss key results of the very recent analysis performed in \cite{FRS-07}, following 
closely the strategy developed in \cite{BFRS}. The starting point is given by 
$B\to\pi\pi$ modes.
Using the $SU(3)$ flavour symmetry of strong interactions and another plausible
dynamical assumption,\footnote{Consistency checks of these working assumptions
can be performed, which are all supported by the current data.} the data for the 
$B\to\pi\pi$ system can be converted into the hadronic parameters of the $B\to\pi K$ 
modes, thereby allowing us to calculate their observables in the SM. Moreover, also 
$\gamma$ can be extracted, with the result
\begin{equation}%\label{}
\gamma=\left(70.0^{+3.8}_{-4.3}\right)^\circ,
\end{equation}
which is in agreement with the SM fits of the UT \cite{CKMfitter,UTfit}.

\begin{figure}
\begin{center}
\includegraphics[width=0.52\textwidth]{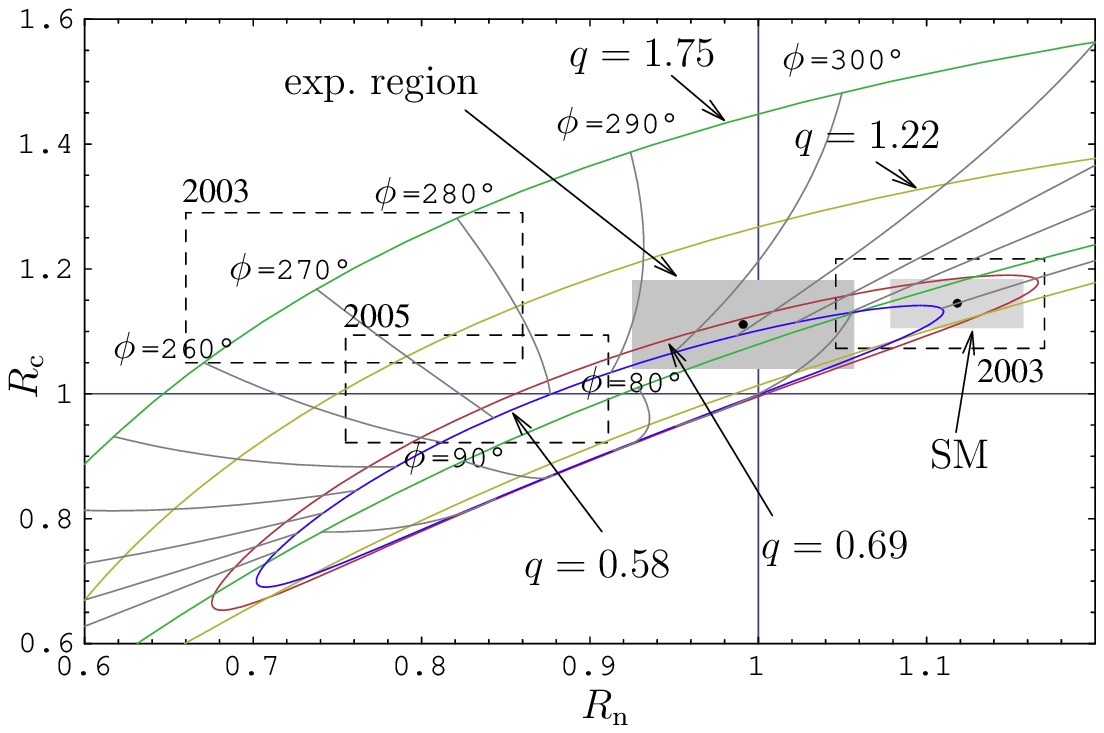}
\end{center}
\vspace*{-0.5truecm}
\caption{\label{fig:RnRc} The situation in the $R_{\rm n}$--$R_{\rm c}$ plane,
as discussed in the text.}
\end{figure}

As far as the $B\to\pi K$ observables with tiny EW penguin contributions are 
concerned, perfect agreement between the SM expectation and the experimental 
data is found. Concerning the $B\to\pi K$ observables receiving sizeable contributions 
from EW penguins, we distinguish between CP-conserving and CP-violating 
observables. In the former case, the key quantities are given by the following 
ratios of CP-averaged $B\to\pi K$ branching ratios \cite{BF98}:
\begin{eqnarray}
R_{\rm c}&\equiv&2\left[\frac{\mbox{BR}(B^+\to\pi^0K^+)+
\mbox{BR}(B^-\to\pi^0K^-)}{\mbox{BR}(B^+\to\pi^+ K^0)+
\mbox{BR}(B^-\to\pi^- \bar K^0)}\right]=1.11\pm0.07\\
R_{\rm n}&\equiv&\frac{1}{2}\left[
\frac{\mbox{BR}(B_d^0\to\pi^- K^+)+\mbox{BR}(\bar B_d^0\to\pi^+ 
K^-)}{\mbox{BR}(B_d^0\to\pi^0K^0)+\mbox{BR}(\bar B_d^0\to\pi^0\bar K^0)}
\right]=0.99\pm0.07,
\end{eqnarray}
where also the most recent experimental averages are indicated \cite{HFAG}.
In these quantities, the EW penguin effects enter in colour-allowed form through 
the modes involving neutral pions, and are theoretically described by a parameter 
$q$, which measures the ``strength" of the EW penguin with respect to the tree 
contributions, and a CP-violating phase $\phi$. In the SM, the $SU(3)$ flavour 
symmetry allows a prediction of $q=0.60$ \cite{NR}, and $\phi$ 
{\it vanishes.} As is known for many years (see, for instance, \cite{EWP-NP}), 
EW penguin topologies offer an interesting avenue for NP to manifest itself in 
the $B$-factory data. In the case of CP-violating NP effects of this kind, $\phi$
would take a value different from zero. 
In Fig.~\ref{fig:RnRc}, we show the situation in the $R_{\rm n}$--$R_{\rm c}$ plane. 
Here the various contours correspond to different values of $q$, and the position on 
the contour is parametrized through the CP-violating phase $\phi$. We observe that 
the SM prediction (on the right-hand side) is very stable in time, having now significantly
reduced errors. On the other hand, the $B$-factory data have moved quite
a bit towards the SM, thereby reducing the ``$B\to\pi K$ puzzle" for the CP-averaged
branching ratios, which emerged already in 2000 \cite{BF00}. In comparison with the situation of the $B\to\pi K$ observables with tiny EW penguin contributions,  
the agreement between the new data for the $R_{\rm c,n}$ and their SM predictions 
is not as perfect. However, a case for a modified EW penguin sector cannot be made 
through the new measurements of these quantities.

\begin{figure}
\vspace*{0.5truecm}
\begin{center}
\includegraphics[width=0.55\textwidth]{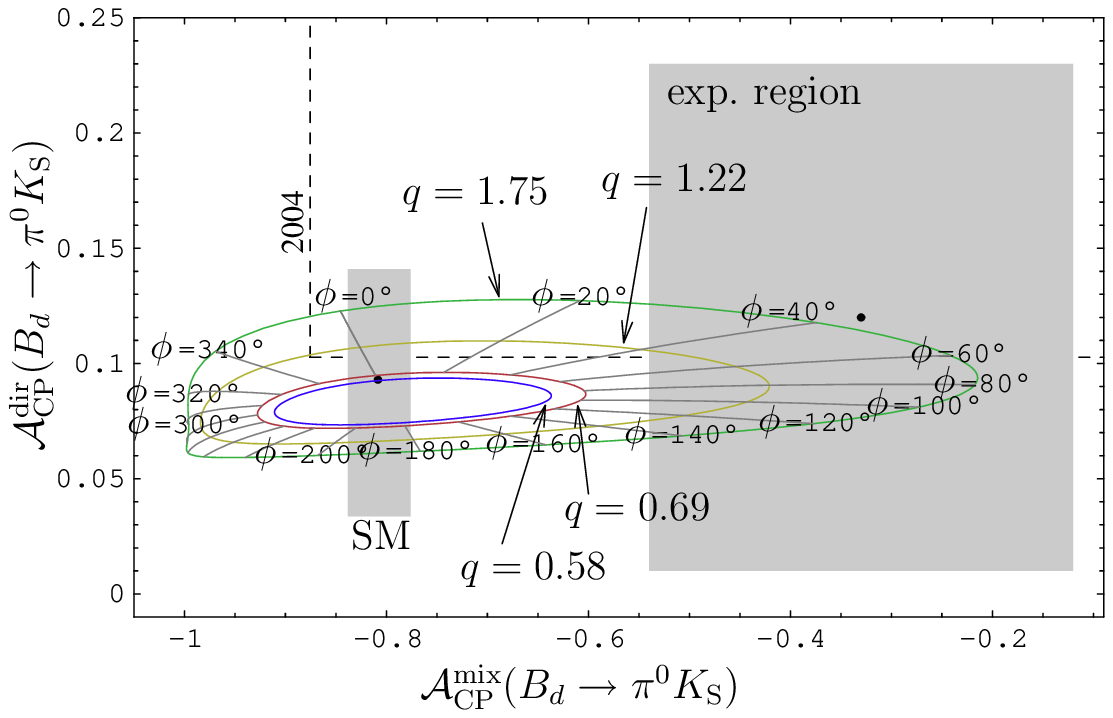}
\end{center}
\vspace*{-0.5truecm}
\caption{\label{fig:ACP}The situation in the 
${\cal A}_{\rm CP}^{\rm mix}(B_d\to\pi^0K_{\rm S})$--${\cal A}_{\rm CP}^{\rm 
dir}(B_d\to\pi^0K_{\rm S})$ plane.}
\end{figure}

Let us now have a closer look at the CP asymmetries of the 
$B^0_d\to\pi^0 K_{\rm S}$ and $B^\pm\to\pi^0K^\pm$ channels. 
As can be seen in Fig.~\ref{fig:ACP}, SM predictions for the CP-violating observables
of $B^0_d\to\pi^0K_{\rm S}$ are obtained that are much sharper than the current
$B$-factory data. In particular ${\cal A}_{\rm CP}^{\rm mix}(B_d\to\pi^0K_{\rm S})$
offers a very interesting quantity. We also see that the experimental central
values can be reached for large {\it positive} values of $\phi$. 
For the new input data, the non-vanishing difference
\begin{equation}\label{Delta-A}
\Delta A \equiv {\cal A}_{\rm CP}^{\rm dir}(B^\pm\to\pi^0K^\pm)-
{\cal A}_{\rm CP}^{\rm dir}(B_d\to\pi^\mp K^\pm)
\stackrel{\rm exp}{=}-0.140\pm0.030
\end{equation}
is likely to be generated through hadronic effects, i.e.\ not through the impact of 
physics beyond the SM. A similar conclusion was drawn in \cite{GR-06}, where 
it was also noted that the measured  values of $R_{\rm c}$ and $R_{\rm n}$ are 
now in accordance with the SM.

Performing, finally, a simultaneous fit to $R_{\rm n}$, $R_{\rm c}$ and the 
CP-violating $B_d\to\pi^0K_{\rm S}$ asymmetries yields
\begin{equation}%\label{}
q=1.7_{-1.3}^{+0.5},\quad \phi=+\left(73_{-18}^{+6}\right)^\circ.
\end{equation}
Interestingly, these parameters -- in particular the large {\it positive} phase -- 
would also allow us to accommodate the experimental values of
$(\sin2\beta)_{\phi K_{\rm S}}$ and the CP asymmetries of other 
$b\to s$ penguin modes with central values smaller than 
$(\sin2\beta)_{\psi K_{\rm S}}$. The large value of $q$ would be excluded 
by constraints from rare decays in simple scenarios where NP enters only 
through $Z$ penguins \cite{BFRS}, but could still be accommodated in other 
scenarios, e.g.\ in models with leptophobic $Z'$ bosons.

\boldmath
\subsection{New Physics in $B^0_d$--$\bar B^0_d$ Mixing}\label{ssec:NP-Bd}
\unboldmath
In the SM, $B^0_d$--$\bar B^0_d$ mixing is governed by box diagrams with internal
top-quark exchances and is, therefore, a strongly suppressed loop phenomenon. In
the presence of NP, we may get new contributions through NP particles in the
box topologies, or new contributions at the tree level (e.g.\ SUSY, $Z'$ models). 
In this case, the off-diagonal element of the mass matrix is modified as 
follows \cite{BF-06}:
\begin{equation}\label{M12d}
M_{12}^{(d)} =M_{12}^{d,{\rm SM}} \left(1 + \kappa_d e^{i\sigma_d}\right),
\end{equation}
where the real parameter $\kappa_d$ is a measure of the strength of NP with 
respect to the SM, and $\sigma_d$ a CP-violating NP phase. The mass difference
$\Delta M_d$ between the two mass eigenstates and the mixing phase $\phi_d$ 
are then modified as
\begin{eqnarray}
\Delta M_d & = &\Delta M_d^{\rm SM}+\Delta M_d^{\rm NP} =
\Delta M_d^{\rm SM}\left| 1 + \kappa_d
  e^{i\sigma_d}\right|\label{DMq-NP}\\
\phi_d & = & \phi_d^{\rm SM}+\phi_d^{\rm NP}=
\phi_d^{\rm SM} + \arg (1+\kappa_d e^{i\sigma_d}),\label{phiq-NP}
\end{eqnarray}
where $\phi_d^{\rm SM}=2\beta$.

\begin{figure}[t]
$$\epsfxsize=0.38\textwidth\epsffile{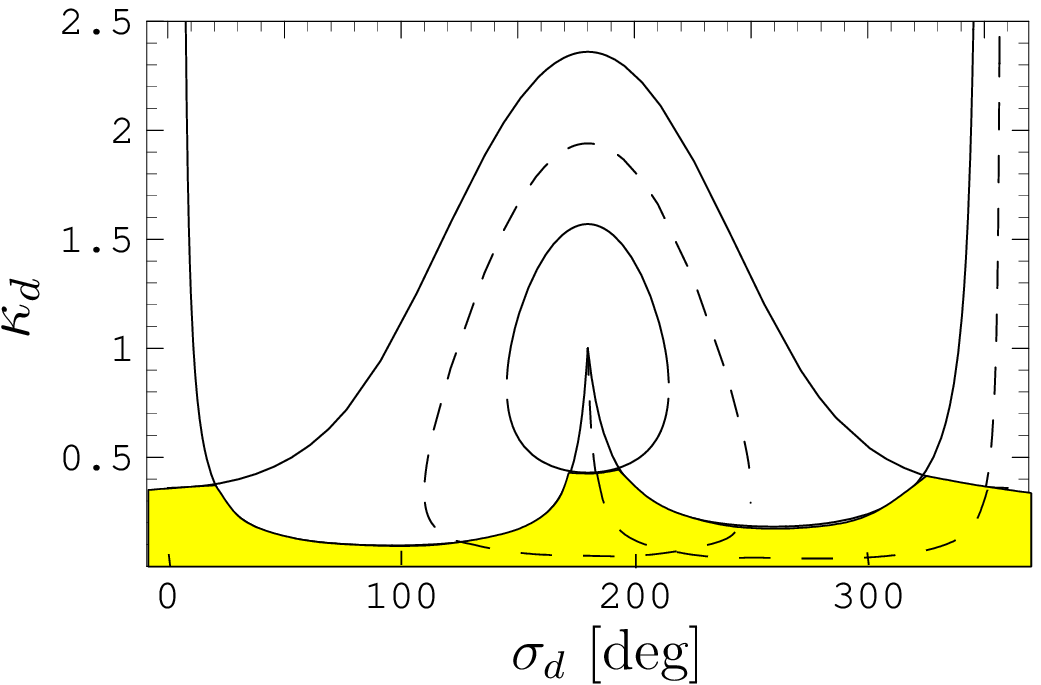}\quad
\epsfxsize=0.38\textwidth\epsffile{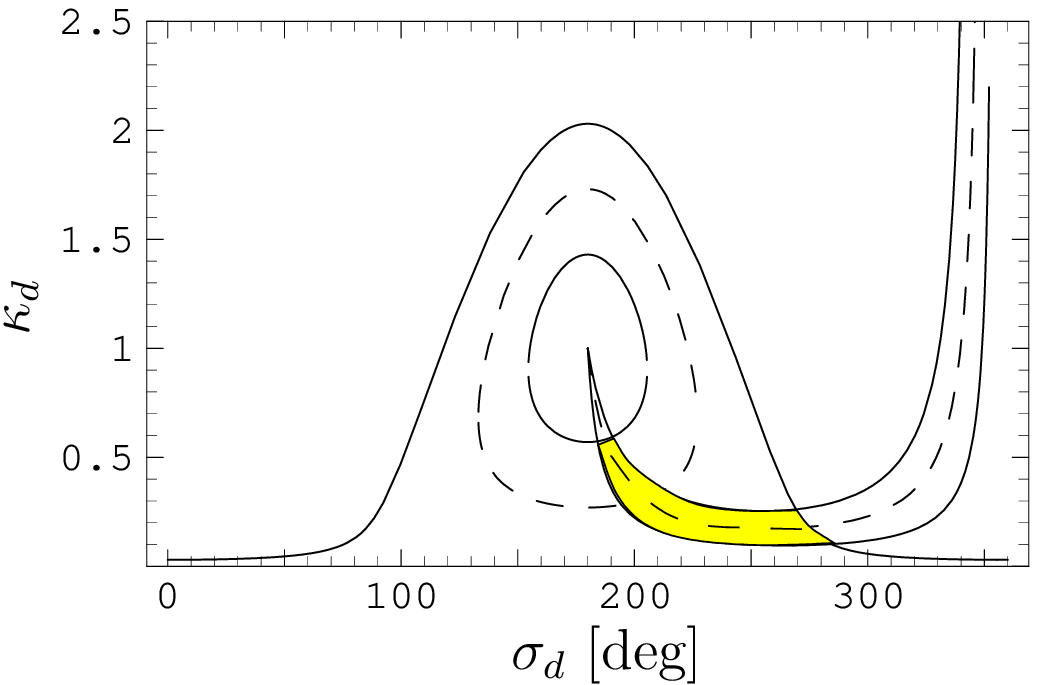}
$$
 \vspace*{-1truecm}
\caption[]{Left panel: allowed region (yellow/grey) in the $\sigma_d$--$\kappa_d$
  plane in a scenario with the JLQCD lattice results and 
  $\phi^{\rm NP}_d|_{\rm excl}$. Right panel: ditto for the 
 scenario with the (HP+JL)QCD   lattice results
 and  $\phi^{\rm NP}_d|_{\rm incl}$. 
}\label{fig:res-k-sig-d}
\end{figure}

Using the $B$-factory data to measure $\Delta M_d$ and to extract the NP 
phase $\phi_d^{\rm NP}$, two sets of contours can be fixed in the 
$\sigma_d$--$\kappa_d$ plane. In the former case, the SM value 
$\Delta M_d^{\rm SM}$ is required. It involves the CKM parameter 
$|V_{td}^*V_{tb}|$, which is governed by $\gamma$ in the corresponding
numerical analysis if the unitarity of the CKM matrix is used. Moreover, information 
about the hadronic parameter $f_{B_d}^2\hat B_{B_d}$ is needed, where 
$f_{B_d}$ is the decay constant of the $B^0_d$ mesons and $B_{B_d}$ the 
``bag" parameter of $B^0_d$--$\bar B^0_d$ mixing, usually coming from lattice
QCD \cite{lubicz}. For the purpose of comparison, 
we use two benchmark sets of such results for these quantities \cite{BF-06}: 
the JLQCD results for two flavours of dynamical light Wilson quarks
\cite{JLQCD}, and a combination of $f_{B_d}$  as determined
by the HPQCD collaboration \cite{HPQCD} for three dynamical flavours with the 
JLQCD result for $\hat B_{B_d}$  [(HP+JL)QCD] \cite{Okamoto}.

For the determination of the NP phase $\phi_d^{\rm NP}=\phi_d-\phi_d^{\rm SM}$, 
we use
\begin{equation}\label{phid-exp}
\phi_d=(42.4\pm2)^\circ,
\end{equation}
which follows from the CP violation in $B_d\to J/\psi K^{(*)}$ decays \cite{HFAG}, 
and fix the ``true" value 
of $\phi_d^{\rm SM}=2\beta$ with the help of the data for tree processes.
This can simply be done through trigonometrical relations between the side 
$R_b\propto |V_{ub}/V_{cb}|$ of the UT and its angle $\gamma$, which are
determined through semileptonic $b\to u \ell \bar\nu_\ell$ decays and 
$B\to D K$ modes, respectively. A numerical analysis shows, that the value of 
$\phi_d^{\rm NP}$ is actually governed by $R_b\propto |V_{ub}/V_{cb}|$, while 
$(\gamma)_{{DK}}$, which suffers currently from large uncertainties \cite{Yamamoto}, 
plays only a minor r\^ole, in contrast to the SM analysis of $\Delta M_d$. 
Unfortunately, we are
facing a discrepancy between the determinations of $|V_{ub}|$ from 
exclusive and inclusive decays \cite{Buchmuller,Mannel}, which has to be resolved in the 
future. The corresponding NP phases read as follows:
\begin{equation}
\phi^{\rm NP}_d|_{\rm excl}=-(3.4\pm7.9)^\circ, \quad 
\phi^{\rm NP}_d|_{\rm incl}=-(11.0\pm4.3)^\circ,
\end{equation} 
where the latter result corresponds to the ``tension" in the fits of the UT
discussed in the context with Fig.~\ref{fig:UT-status}. The resulting situation in 
the $\sigma_d$--$\kappa_d$ plane is shown in Fig.~\ref{fig:res-k-sig-d}, where 
the hill-like structures correspond to the constraints from $\Delta M_d$, which
are complementary to those of $\phi^{\rm NP}_d$. We observe that the measurement
of CP violation in $B_d\to J/\psi K^{(*)}$ decays has a dramatic impact on the
allowed region in NP parameter space; the right panel may indicate the presence 
of NP, although no definite conclusions can be drawn at the moment. It will
be interesting to monitor this effect in the future. In order to detect such CP-violating
NP contributions, things are much more promising in the $B_s$ system.

\boldmath
\section{Key Targets of $B$-Physics Studies at the LHC}
\unboldmath
The exploration of $B$-meson decays at hadron colliders -- and the LHC in 
particular -- is characterized through a high statistics and the access
the $B_s$-meson system, which offers a physics programme that is
to a large extent complementary to that of the $e^+e^-$ $B$ factories 
operating at the $\Upsilon(4S)$ resonance. 

\boldmath
\subsection{General Features of the $B_s$ System}
\unboldmath
For $B^0_s$-mesons, we expect -- within the SM -- a mass difference 
$\Delta M_s={\cal O}(20\,\mbox{ps}^{-1})$, which is much larger than the
experimental value of $\Delta M_d = 0.5\,\mbox{ps}^{-1}$. Consequently,
the $B^0_s$--$\bar B^0_s$ oscillations are very rapid, thereby making it
very challenging to resolve them experimentally. 

Whereas the difference between the decay widths of the mass eigenstates
of the $B^0_d$-meson system is negligible, its counterpart 
$\Delta\Gamma_s/\Gamma_s$ in the $B^0_s$-meson system is expected to 
be of ${\cal O}(10\%)$ \cite{lenz}. Recently, the first results for 
$\Delta\Gamma_s$ were reported from the Tevatron, using the 
$B^0_s\to J/\psi\phi$ channel \cite{DDF}:
\begin{equation}\label{DG-det}
\frac{\Delta\Gamma_s}{\Gamma_s}=\left\{
\begin{array}{ll}
0.65^{+0.25}_{-0.33}\pm0.01 & \mbox{(CDF \cite{CDF-DG})}\\
0.24^{+0.28+0.03}_{-0.38-0.04} & \mbox{(D0 \cite{D0-DG})}.
\end{array}
\right.
\end{equation}
It will be interesting to follow the evolution of the data for this quantity; at the
LHC, we expect a precision of about $0.01$  after one year of
taking data \cite{schopper,nakada}. The width difference 
$\Delta\Gamma_s$ offers studies of CP violation through ``untagged" rates 
of the following form:
\begin{equation}%\label{}
\langle\Gamma(B_s(t)\to f)\rangle
\equiv\Gamma(B^0_s(t)\to f)+\Gamma(\overline{B^0_s}(t)\to f),
\end{equation}
which are interesting in terms of efficiency, acceptance and purity. If 
both $B^0_s$ and $\bar B^0_s$ states may decay into the final state $f$,
the rapidly oscillating $\Delta M_st$ terms cancel. Various ``untagged" 
strategies exploiting this feature were proposed (see \cite{DDF} and 
\cite{dun}--\cite{DFN}); we will discuss an example in Section~\ref{ssec:Bspsiphi}.

Finally, the CP-violating phase of $B^0_s$--$\bar B^0_s$ mixing is tiny in the SM:
\begin{equation}\label{phis-SM}
\phi_s^{\rm SM}=-2\lambda^2\eta\approx -2^\circ,
\end{equation}
which is very interesting for the search of signals of NP \cite{DFN,NiSi,BMPR}
(see Section~\ref{ssec:Bspsiphi}).

\begin{figure}[t]
$$\epsfxsize=0.40\textwidth\epsffile{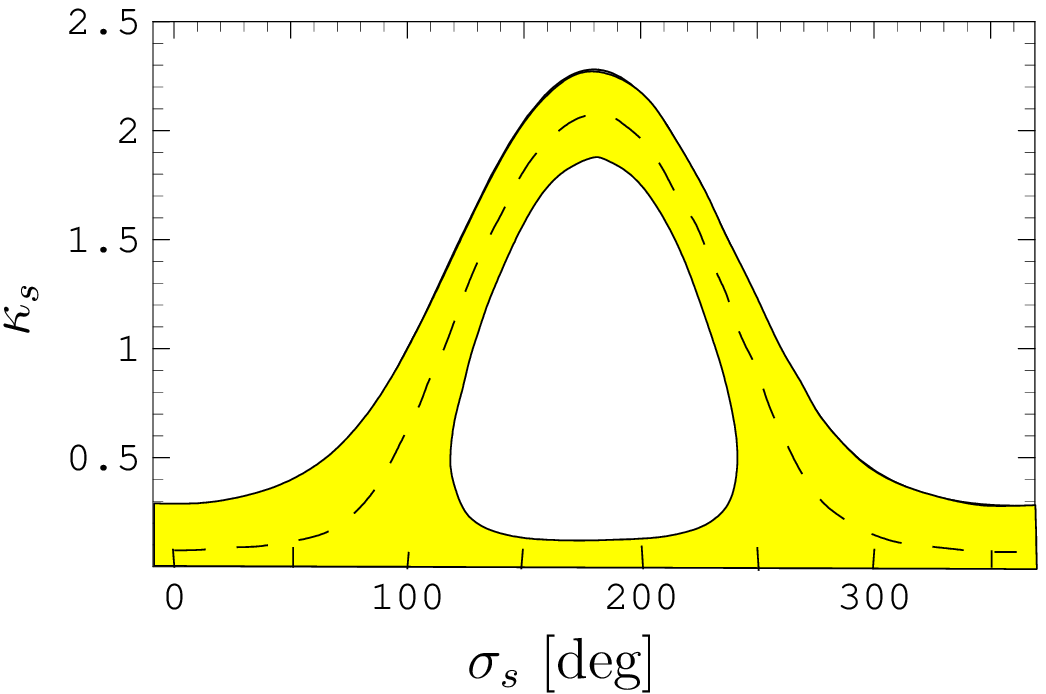}\quad
\epsfxsize=0.40\textwidth\epsffile{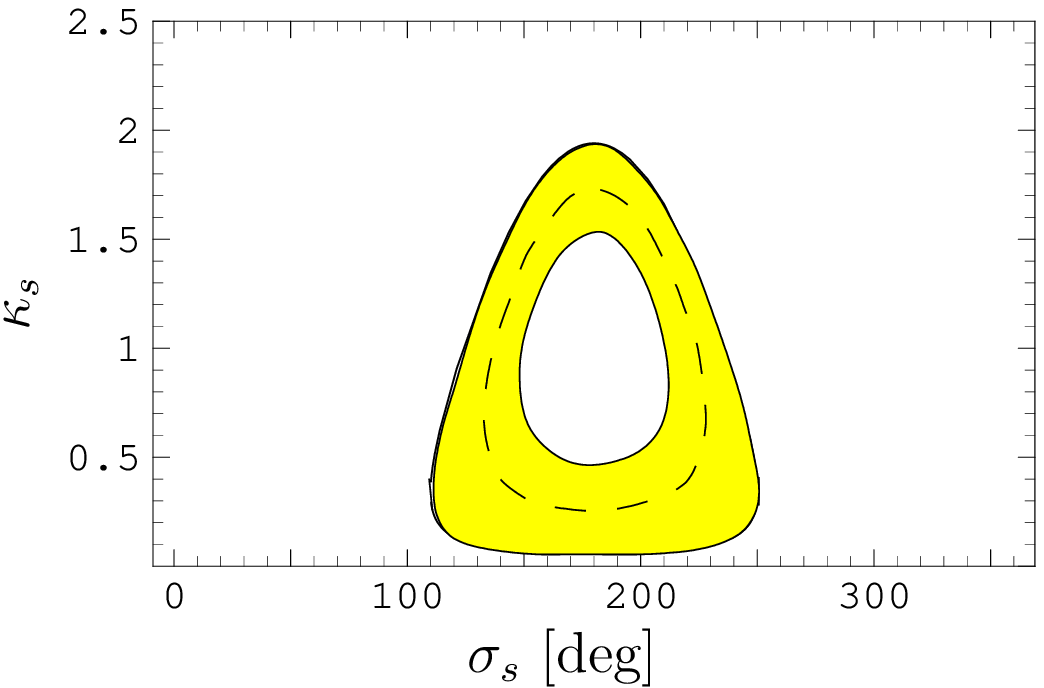}
$$
 \vspace*{-1truecm}
\caption[]{The allowed regions (yellow/grey) in the $\sigma_s$--$\kappa_s$ plane.
Left panel: JLQCD lattice results. Right panel: (HP+JL)QCD lattice 
results.}\label{fig:MDs-NP}
\end{figure}

\boldmath
\subsection{Hot News of 2006: Measurement of $\Delta M_s$}
\unboldmath
For many years, only lower bounds on $\Delta M_s$ were available from
the LEP (CERN) experiments and SLD (SLAC) \cite{Bosc-WG}. In 2006, the value 
of $\Delta M_s$ could eventually be pinned down at the Tevatron \cite{menzemer}:
the D0 collaboration reported a two-sided bound 
\begin{equation}\label{D0-range}
17 \,{\rm ps}^{-1}< \Delta M_s < 21\,{\rm ps}^{-1} \quad \mbox{(90\% C.L.)},
\end{equation}
corresponding to a 2.5\,$\sigma$ signal at $\Delta M_s=19\,{\rm ps}^{-1}$ 
\cite{D0}, and CDF announced the following result \cite{CDF}:
\begin{equation}\label{CDF-DMs}
\Delta M_s = \left[17.77\pm0.10({\rm stat})
\pm 0.07({\rm syst})\right]\,{\rm ps}^{-1},
\end{equation}
which corresponds to a signal at the $5\,\sigma$ level. These new experimental
results have immediately triggered a lot of theoretical activity (see, e.g.,
\cite{DMs-papers,BBGT}). 

Let us here follow once again the analysis performed in \cite{BF-06}. In order to explore
the allowed region in NP parameter space that follows from the measurements
at the Tevatron, we have just to make the substitution $d\to s$ in (\ref{M12d}).
Using the unitarity of the CKM matrix and the Wolfenstein expansion, the
CKM factor entering the SM expression for $\Delta M_s$ takes the simple form
\begin{equation}%\label{}
|V_{ts}^*V_{tb}|=|V_{cb}|\left[1+{\cal O}(\lambda^2)\right].
\end{equation}
Consequently, in contrast to the SM analysis of $\Delta M_d$, no information 
on $\gamma$ and $R_b$ is needed in this expression, which is an important
advantage. The accuracy of the SM prediction of $\Delta M_s$ is hence  limited 
by the hadronic parameter $f_{B_s}\hat{B}_{B_s}^{1/2}$. 
Recently, the HPQCD collaboration has reported the result 
$\Delta M_s^{\rm SM}=20.3(3.0)(0.8)\,\mbox{ps}^{-1}$ \cite{HPQCD-DMs}, 
which lies between the $\Delta M_s^{\rm SM}|_{\rm JLQCD}=(16.1\pm 2.8) 
\,{\rm ps}^{-1}$ and $\Delta M_s^{\rm SM}|_{\rm (HP+JL)QCD}  =  
(23.4\pm 3.8)\,{\rm ps}^{-1}$ results entering Fig.~\ref{fig:MDs-NP}. 
In this figure, which corresponds to Fig.~\ref{fig:res-k-sig-d},
we show the allowed regions in the $\sigma_s$--$\kappa_s$ plane.
We see that  the measurement of $\Delta M_s$ leaves ample space for the 
NP parameters $\sigma_s$ and $\kappa_s$. The experimental errors are 
already significantly smaller than the theoretical ones. Any more precise statement 
about the presence or absence of NP in the mass difference $\Delta M_s$ requires 
the reduction of the theoretical lattice QCD uncertainties. 

As discussed in \cite{BF-06}, the situation is not much better for constraints
on NP through $\Delta M_s/\Delta M_d$. In the analysis of this ratio 
an $SU(3)$-breaking parameter 
\begin{equation}%\label{}
\xi \equiv \frac{f_{B_s}\hat{B}_{B_s}^{1/2}}{f_{B_d}\hat{B}_{B_d}^{1/2}}
\end{equation}
enters, which has a reduced theoretical uncertainty in comparison with 
the individual values of the $f_{B_q}\hat B_{B_q}^{1/2}$. Usually, 
$\Delta M_s/\Delta M_d$ is used for the determination of the side 
$R_t\propto |V_{td}/V_{cb}|=|V_{td}/V_{ts}|\left[1+{\cal O}(\lambda^2)\right]$
of the UT. Alternatively, applying the unitarity of the CKM matrix, the
following quantity can be determined:
\begin{equation}\label{DMsDMs-ratio}
\frac{\rho_s}{\rho_d}=\lambda^2
\underbrace{\left[1-2R_b\cos\gamma+R_b^2\right]}_{=R_t^2}
\left[1+%(1-2R_b\cos\gamma)\lambda^2+
{\cal O}(\lambda^2)\right]
\frac{1}{\xi^2}\frac{M_{B_d}}{M_{B_s}}
\frac{\Delta M_s}{\Delta M_d},
\end{equation}
where the ratio on the left-hand side equals 1 in the SM. For the current data,
$\gamma$ is the major source of uncertainty, in addition to the hadronic
parameter $\xi$. Thanks to precision measurements of $\gamma$ at LHCb,
the CKM and lattice uncertainties should be of the same order of magnitude
by 2010. However, unless the central values move dramatically, we would
still get a result in agreement with 1 \cite{BF-06}. This case could correspond
to the SM, but could also have NP contributions that enter in the same manner
in $\Delta M_s$ and $\Delta M_d$. Consequently, we would still be left with a
rather unsatisfactory situation concerning the search for signals of NP
through (\ref{DMsDMs-ratio}), even after a couple of years taking data at LHCb.

As in the case of the $B_d$-meson system discussed in Section~\ref{ssec:NP-Bd}, 
the allowed region in the $\sigma_s$--$\kappa_s$ plane will be dramatically
reduced as soon as measurements of CP violation in the $B_s$-meson 
system become available. The ``golden" channel in this respect is given 
by $B^0_s\to J/\psi \phi$, which is our next topic.

\boldmath
\subsection{The Decay $B^0_s\to J/\psi \phi$}\label{ssec:Bspsiphi}
\unboldmath
This mode is the counterpart of the $B^0_d\to J/\psi K_{\rm S}$ transition, where
we have just to replace the down quark by a strange quark. The structures of the
corresponding decay amplitudes are completely analogous to each other. However,
there is also an important difference with respect to $B^0_d\to J/\psi K_{\rm S}$,
since the final state of $B^0_s\to J/\psi \phi$ contains two vector mesons and is,
hence, an admixture of different CP eigenstates. Using the angular distribution of the 
$J/\psi [\to\ell^+\ell^-]\phi [\to\ K^+K^-]$ decay products, the CP eigenstates
can be disentangled \cite{DDLR} and the time-dependent decay rates calculated
\cite{DDF,DFN}. As in the case of $B^0_d\to J/\psi K_{\rm S}$, the
hadronic matrix elements cancel then in the mixing-induced observables. For the
practical implementation, a set of three linear polarization amplitudes is usually 
used: $A_0(t)$ and $A_\parallel(t)$ correspond to CP-even final-state configurations,
whereas $A_\perp(t)$ describes a CP-odd final-state configuration.

It is instructive to illustrate how this works by having a closer look at the
one-angle distribution, which takes the following form \cite{DDF,DFN}:
\begin{equation}%\label{}
\frac{d\Gamma(B^0_s(t)\to J/\psi \phi)}{d\cos\Theta}\propto
\left(|A_0(t)|^2+|A_\parallel(t)|^2\right)
\frac{3}{8}\left(1+\cos^2\Theta\right)+|A_\perp(t)|^2\frac{3}{4}\sin^2\Theta.
\end{equation}
Here $\Theta$ is defined as the angle between the momentum of the $\ell^+$
and the normal to the decay plane of the $K^+K^-$ system in the $J/\psi$
rest frame. The time-dependent measurement of the angular dependence
allows us to extract the following observables:
\begin{equation}%\label{}
P_+(t)\equiv |A_0(t)|^2+|A_\parallel(t)|^2, \quad
P_-(t)\equiv |A_\perp(t)|^2,
\end{equation}
where $P_+(t)$ and $P_-(t)$ refer to the CP-even and CP-odd final-state configurations,
respectively. If we consider the case of having an initially, i.e.\ at time $t=0$, present
$\bar B^0_s$ meson, the CP-conjugate quantities $\bar P_\pm(t)$ can be extracted
as well. Using an {\it untagged} data sample, the untagged rates
\begin{equation}%\label{}
P_\pm(t)+\overline{P}_\pm(t)\propto
\left[(1\pm\cos\phi_s)e^{-\Gamma_{\rm L}t}+
(1\mp\cos\phi_s)e^{-\Gamma_{\rm H}t}\right]
\end{equation}
can be determined, while a {\it tagged} data sample allows us to measure
the CP-violating asymmetries
\begin{equation}%\label{}
\frac{P_\pm(t)-\overline{P}_\pm(t)}{P_\pm(t)+\overline{P}_\pm(t)}=
\pm\left[\frac{2\,\sin(\Delta M_st)\sin\phi_s}{(1\pm\cos\phi_s)e^{+\Delta\Gamma_st/2}+
(1\mp\cos\phi_s)e^{-\Delta\Gamma_st/2}}\right].
\end{equation}
In the presence of CP-violating NP contributions to $B^0_s$--$\bar B^0_s$
mixing, we obtain
\begin{equation}%\label{}
\phi_s=-2\lambda^2R_b\sin\gamma+\phi_s^{\rm NP}\approx -2^\circ+
\phi_s^{\rm NP}\approx \phi_s^{\rm NP}.
\end{equation}
Consequently, NP of this kind would be indicated by the following features:
\begin{itemize}
\item The {\it untagged} observables depend on {\it two} exponentials;
\item {\it sizeable} values of the CP-violating asymmetries.
\end{itemize}
It should be emphasized that this avenue to search for NP signals does {\it not}
have to rely on lattice QCD results, in contrast the analyses of $\Delta M_s$
discussed above.

These general features hold also for the full three-angle distribution
\cite{DDF,DFN}: it is much more involved than the one-angle case, but
provides also additional information through interference terms of the
form 
\begin{equation}
\mbox{Re}\{A_0^\ast(t)A_\parallel(t)\}, \quad
\mbox{Im}\{A_f^\ast(t)A_\perp(t)\} \, (f\in\{0,\parallel\}).
\end{equation}
From an experimental point of view, there is no experimental draw-back with
respect to the one-angle case. Following these lines, $\Delta\Gamma_s$ 
(see (\ref{DG-det})) and $\phi_s$ can be extracted.  
Recently, the D0 collaboration has
reported first results for the measurement of $\phi_s$ through the
untagged, time-dependent three-angle $B^0_s\to J/\psi\phi$ distribution
\cite{D0-phis}:
\begin{equation}%\label{}
\phi_s=-0.79\pm0.56\,\mbox{(stat.)} ^{+0.14}_{-0.01}\,\mbox{(syst.)}
=-(45\pm32^{+1}_{-8})^\circ.
\end{equation}
This phase is therefore not yet stringently constrained. However, it will be
very accessible at the LHC, where the following picture is expected with 
nominal one year data \cite{nakada}: if $\phi_s$ takes its SM value,
a $2\,\sigma$ measurement will be possible at LHCb (2\,$\mbox{fb}^{-1}$),
ATLAS and CMS expect uncertainties of ${\cal O}(0.1)$ (10\,$\mbox{fb}^{-1}$) 
\cite{smsp}. At some point, also in view of LHCb upgrade plans \cite{LHCb-up}, 
we have to include hadronic penguin uncertainties. This can 
be done with the help of the $B^0_d\to J/\psi \rho^0$ decay \cite{RF-ang}.

\begin{figure}[t] 
$$\epsfxsize=0.40\textwidth\epsffile{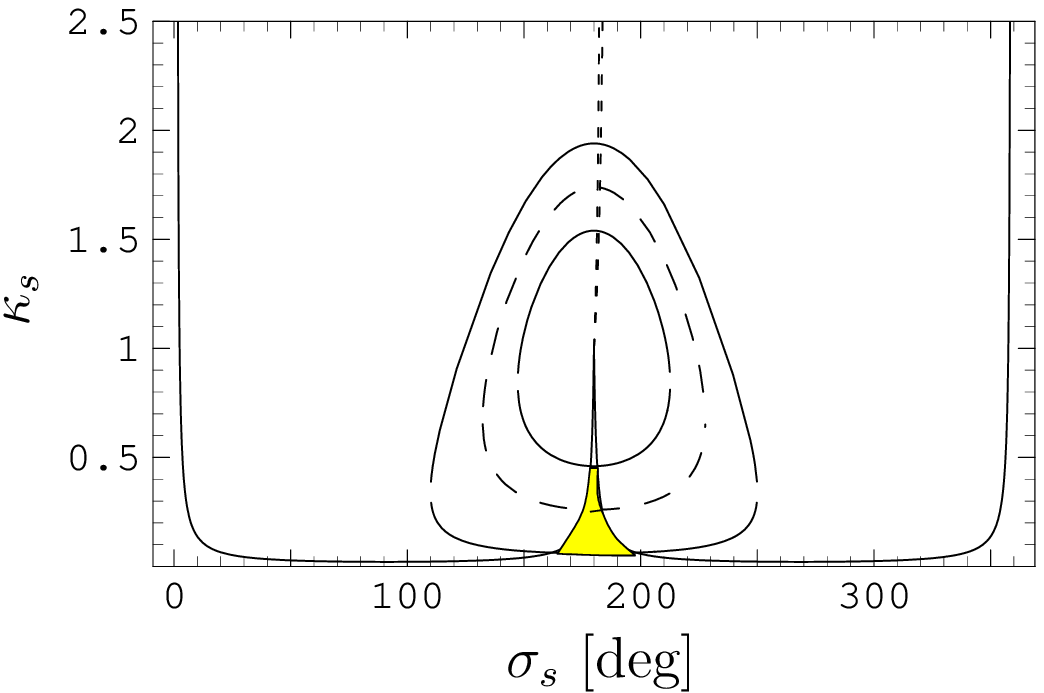}\quad
\epsfxsize=0.40\textwidth\epsffile{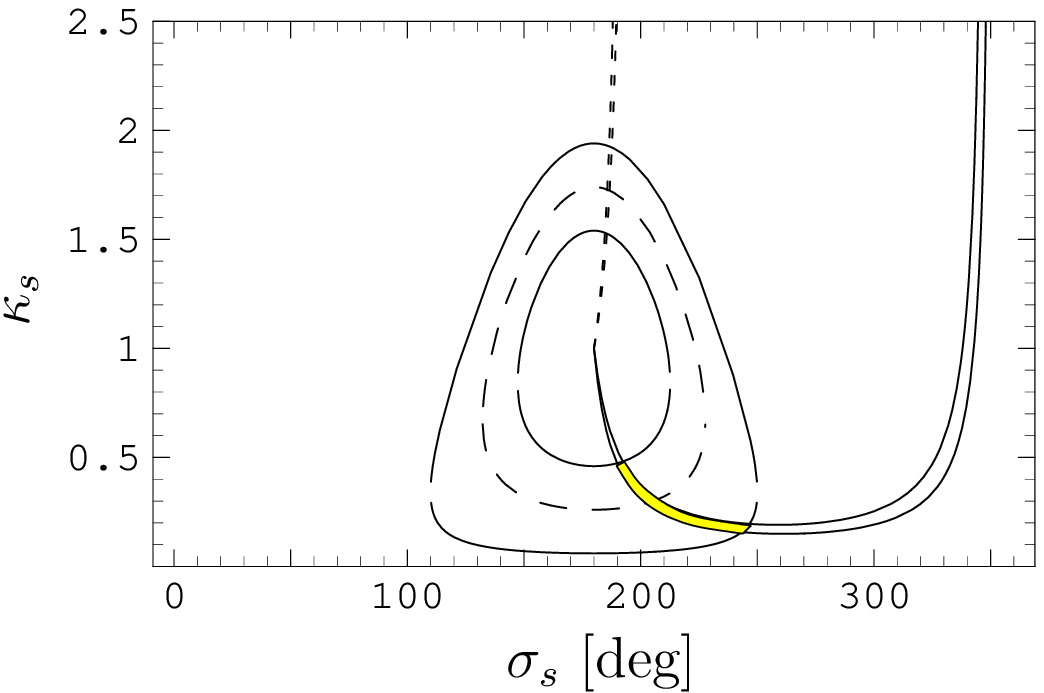}$$
\vspace*{-1cm}
   \caption[]{Illustration of the impact of measurements of CP violation in 
   $B^0_s\to J/\psi\phi$ for the two 2010 scenarios i) [left panel] and ii) [right panel]
   discussed in the text.}\label{fig:sis-kas-CP}
\end{figure}

In order to illustrate the impact of measurements of CP violation in $B^0_s\to J/\psi\phi$,
let us discuss two scenarios for the year 2010 \cite{BF-06}:
\begin{itemize}
\item[i)] $(\sin\phi_s)_{\rm exp}=-0.04\pm0.02$: this case corresponds to the SM;
\item[ii)] $(\sin\phi_s)_{\rm exp}=-0.20\pm0.02$: such a measurement would
give a NP signal at the $10\,\sigma$ level. This scenario corresponds to a simple
translation of the ``tension" in the UT fits discussed above for $\kappa_s=\kappa_d$, 
$\sigma_s=\sigma_d$, and demonstrates the power of the $B_s$
system to search for NP.  
\end{itemize}
We see that it will be very challenging to establish NP effects in $B^0_s$--$\bar B^0_s$
mixing without new CP-violating contributions to this phenomenon. However, the
data still leave a lot of space for such effects in specific scenarios (e.g.\ SUSY,
extra $Z'$ and little Higgs models \cite{BF-06,DMs-papers,LH}), 
which could be detected at the LHC. It will be very exciting to follow the 
corresponding measurements after the start of this new collider.

\subsection{Further Benchmark Decays for LHCb}
This experiment has a very rich physics programme (for an experimental 
overview, see \cite{schopper}). Besides many other interesting aspects, there 
are two major lines of research:
\begin{enumerate}
\item Precision measurements of $\gamma$:\\
On the one hand, there are strategies using pure tree decays: 
$B^0_s\to D_s^\mp K^\pm$ [$\sigma_\gamma\sim14^\circ$],
$B^0_d\to D^0K^{*}$ [$\sigma_\gamma\sim8^\circ$],
$B^\pm\to D^0K^\pm$ [$\sigma_\gamma\sim5^\circ$],
where we have also indicated the expected sensitivities after 
one year of taking data \cite{schopper}. These numbers should be
compared with the current $B$-factory data, yielding
\begin{equation}%\label{}
\left.\gamma\right|_{D^{(*)} K^{(*)}} = \left\{
\begin{array}{ll}
(62^{+38}_{-24})^\circ & \mbox{(CKMfitter)}\\[5pt] 
(82\pm 20)^\circ & \mbox{(UTfit).}
\end{array}
\right.
\end{equation}
These extractions are very robust with respect to NP effects.
On the other hand, $\gamma$ can also be extracted from $B$-meson decays
with penguin contributions: $B^0_s\to K^+K^-$ and $B^0_d\to \pi^+\pi^-$
[$\sigma_\gamma\sim5^\circ$], $B^0_s\to D_s^+D_s^-$ and $B^0_d\to D_d^+D_d^-$.
The key question is whether discrepancies will arise in these determinations. 

\item Analyses of rare decays, which are absent at the SM tree level:\\ 
prominent examples are $B^0_{s,d}\to\mu^+\mu^-$,
$B^0_d\to K^{*0}\mu^+\mu^-$ and $B^0_s\to \phi \mu^+\mu^-$. In order to 
complement the studies of $B^0_d\to \phi K_{\rm S}$ at the $B$ factories discussed
in Section~\ref{ssec:bsss}, $B^0_s\to\phi\phi$ is a very interesting mode for LHCb.  
\end{enumerate}
Let us next have a closer look at some of these decays.

\boldmath
\subsubsection{CP Violation in $B_s\to D_s^\pm K^\mp$ and $B_d\to D^\pm\pi^\mp$}
\unboldmath
The pure tree decays $B_s\to D_s^\pm K^\mp$ \cite{BsDsK} and 
$B_d\to D^\pm \pi^\mp$ \cite{BdDpi} can be treated on the same theoretical 
basis, and provide new strategies to determine $\gamma$ \cite{RF-gam-ca}. 
Following this paper, we write these modes as $B_q\to D_q \bar u_q$. Their
characteristic feature is that both a $B^0_q$ and a $\bar B^0_q$ meson may decay 
into the same final state $D_q \bar u_q$. Consequently,  interference effects 
between $B^0_q$--$\bar B^0_q$ mixing and decay processes arise, which 
involve the CP-violating phase combination $\phi_q+\gamma$.

In the case of $q=s$, i.e.\ $D_s\in\{D_s^+, D_s^{\ast+}, ...\}$ and 
$u_s\in\{K^+, K^{\ast+}, ...\}$, these interference effects are governed 
by a hadronic parameter $X_s e^{i\delta_s}\propto R_b\approx0.4$, where
$R_b\propto |V_{ub}/V_{cb}|$ is the usual UT side, and hence are large. 
On the other hand, for $q=d$, i.e.\ $D_d\in\{D^+, D^{\ast+}, ...\}$ 
and $u_d\in\{\pi^+, \rho^+, ...\}$, the interference effects are described 
by $X_d e^{i\delta_d}\propto -\lambda^2R_b\approx-0.02$, and hence are tiny. 

Measuring the $\cos(\Delta M_qt)$ and $\sin(\Delta M_qt)$ terms of the
time-dependent $B_q\to D_q \bar u_q$ rates, a theoretically clean 
determination of $\phi_q+\gamma$ is possible \cite{BsDsK,BdDpi}. 
Since the $\phi_q$ can be determined separately, $\gamma$ 
can be extracted. However, in the practical implementation, there are problems:
we encounter an eightfold discrete ambiguity for $\phi_q+\gamma$, which
is very disturbing for the search of NP, and in the $q=d$ case, an additional input 
is required to extract $X_d$ since
${\cal O}(X_d^2)$ interference effects would otherwise have to be resolved,
which is impossible. Performing a combined analysis of the $B^0_s\to D_s^{+}K^-$
and $B^0_d\to D^+\pi^-$ decays, these problems can be solved \cite{RF-gam-ca}.
This strategy exploits the fact that these transitions are related to each other
through the $U$-spin symmetry of strong interactions,\footnote{The $U$ spin is an
$SU(2)$ subgroup of the $SU(3)_{\rm F}$ flavour-symmetry group of QCD, connecting
$d$ and $s$ quarks in analogy to the isospin symmetry, which relates $d$ and
$u$ quarks to each other.} allowing us to simplify the hadronic sector. Following
these lines, an unambiguous value of $\gamma$ can be extracted from the
observables. To this end, $X_d$ has actually not to be fixed, and $X_s$ may only enter
through a $1+X_s^2$ correction, which is determined through untagged $B_s$
rates. The first studies for LHCb are very promising \cite{WG5-rep}, and are 
currently further refined.

\boldmath
\subsubsection{The $B_s\to K^+K^-$, $B_d\to \pi^+\pi^-$ System}
\begin{figure}[t]
   \centering
  \includegraphics[width=7.0truecm]{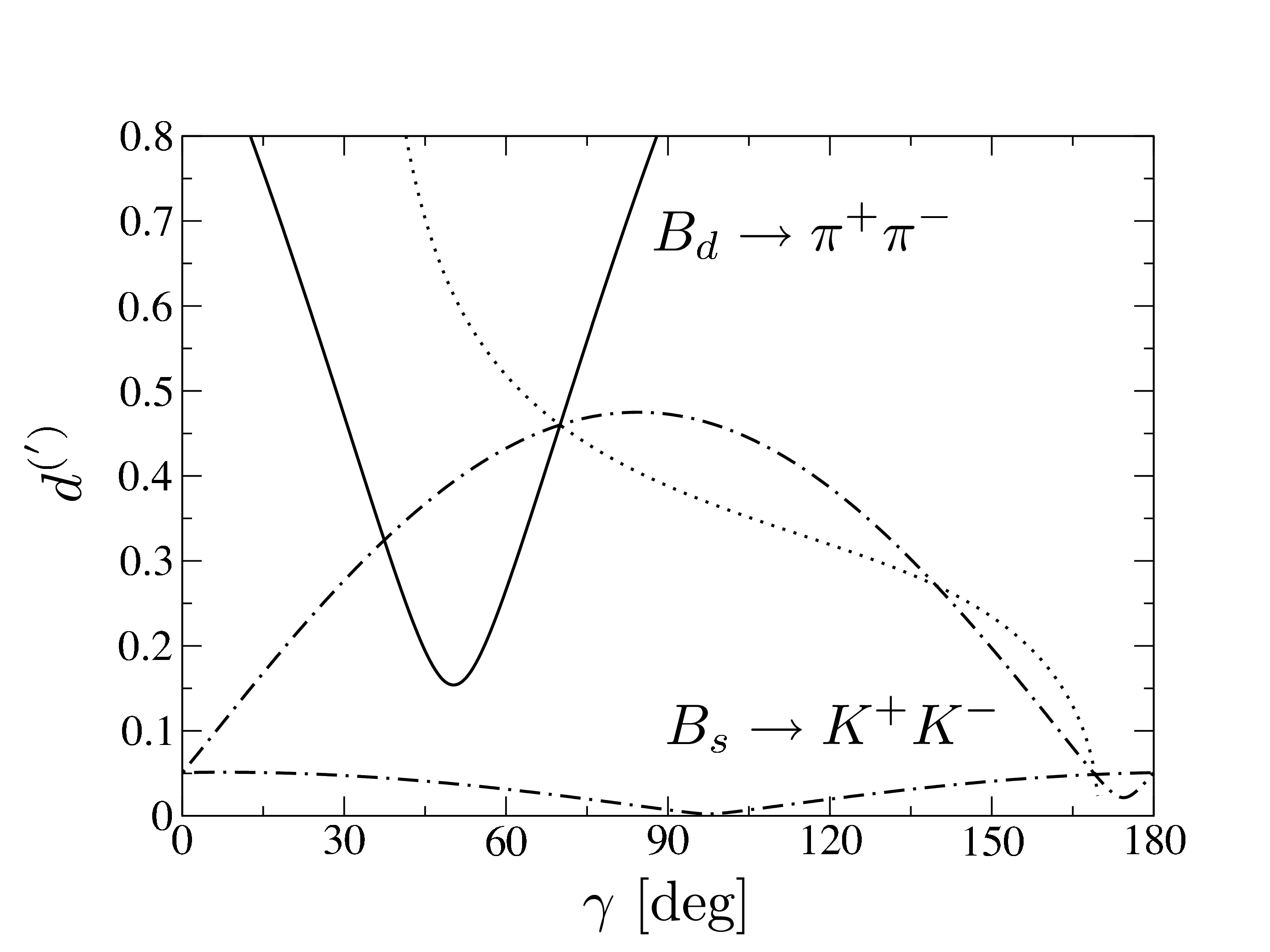} 
     \caption{The contours in the $\gamma$--$d^{(')}$ plane for an example with
   $d=d'=0.46$, $\theta=\theta'=155^\circ$, $\phi_d=42.4^\circ$, $\phi_s=-2^\circ$,
   $\gamma=70^\circ$, which corresponds to the CP asymmetries
   ${\cal A}_{\rm CP}^{\rm dir}(B_d\to\pi^+\pi^-)=-0.24$ and 
   ${\cal A}_{\rm CP}^{\rm mix}(B_d\to\pi^+\pi^-)=+0.59$, as well as
   ${\cal A}_{\rm CP}^{\rm dir}(B_s\to K^+K^-)=+0.09$ and
   ${\cal A}_{\rm CP}^{\rm mix}(B_s\to K^+K^-)=-0.23$.}\label{fig:Bs-Bd-contours}
\end{figure}

\unboldmath
The decay $B^0_s\to K^+K^-$ is a $\bar b \to \bar s$ transition, and
involves tree and penguin amplitudes, as $B^0_d\to\pi^+\pi^-$ \cite{RF-BsKK}. 
However, because of the different CKM structure, the latter 
topologies play actually the dominant r\^ole in $B^0_s\to K^+K^-$,
whereas the major contribution to $B^0_d\to\pi^+\pi^-$ is due to the tree
amplitude.  In the SM, we may write
\begin{eqnarray}
A(B^0_d\to\pi^+\pi^-)& \propto & \left[e^{i\gamma}-de^{i\theta}\right]\\
A(B_s^0\to K^+K^-)& \propto &
\left[e^{i\gamma}+\left(\frac{1-\lambda^2}{\lambda^2}\right)d'e^{i\theta'}\right],
\end{eqnarray}
where the CP-conserving hadronic parameters $de^{i\theta}$ and
$d'e^{i\theta'}$ descripe -- sloppily speaking -- the ratios of penguin to tree
contributions. The direct and mixing-induced CP asymmetries take then the 
following general form:
\begin{equation}
{\cal A}_{\rm CP}^{\rm dir}(B_d\to \pi^+\pi^-)=
G_1(d,\theta;\gamma), \quad
{\cal A}_{\rm CP}^{\rm mix}(B_d\to \pi^+\pi^-)=
G_2(d,\theta;\gamma,\phi_d)
\end{equation}
\begin{equation}
{\cal A}_{\rm CP}^{\rm dir}(B_s\to K^+K^-)=
G_1'(d',\theta';\gamma), \quad
{\cal A}_{\rm CP}^{\rm mix}(B_s\to K^+K^-)=
G_2'(d',\theta';\gamma,\phi_s).
\end{equation}
Since $\phi_d$ is already known (see (\ref{phid-exp})) and $\phi_s$ is negligibly small
in the SM -- or can be determined through $B^0_s\to J/\psi \phi$ should CP-violating
NP contributions to $B^0_s$--$\bar B^0_s$ mixing make it sizeable -- 
we may convert the measured values of 
${\cal A}_{\rm CP}^{\rm dir}(B_d\to \pi^+\pi^-)$, 
${\cal A}_{\rm CP}^{\rm mix}(B_d\to \pi^+\pi^-)$ and
${\cal A}_{\rm CP}^{\rm dir}(B_s\to K^+K^-)$, 
${\cal A}_{\rm CP}^{\rm mix}(B_s\to K^+K^-)$ into {\it theoretically clean}
contours in the $\gamma$--$d$ and $\gamma$--$d'$ planes, respectively.
In Fig.~\ref{fig:Bs-Bd-contours}, we show these contours (solid and dot-dashed) 
for an example, which is inspired by the current $B$-factory data \cite{FRS-07}.

A closer look at the corresponding Feynman diagrams shows that 
$B^0_d\to\pi^+\pi^-$ is actually related to $B^0_s\to K^+K^-$ through the interchange 
of all down and strange quarks. Consequently, each decay topology contributing
to $B^0_d\to\pi^+\pi^-$ has a counterpart in $B^0_s\to K^+K^-$ and vice versa, 
and the corresponding hadronic parameters can be related to each other
with the help of the $U$-spin flavour symmetry of strong interactions,
implying the following relations \cite{RF-BsKK}:
\begin{equation}\label{U-spin-rel}
d'=d, \quad \theta'=\theta.
\end{equation}
Applying the former, we may extract $\gamma$ and $d$ through the 
intersections of the theoretically clean $\gamma$--$d$ and $\gamma$--$d'$ 
contours. In the example of Fig.~\ref{fig:Bs-Bd-contours}, a twofold
ambiguity arises from the solid and dot-dashed curves. However, as 
discussed in \cite{RF-BsKK}, it can be resolved with the help of the dotted 
contour, thereby leaving us with the ``true" solution of $\gamma=70^\circ$ in this 
case. Moreover, we may determine $\theta$ and $\theta'$, which allow an interesting 
internal consistency check of the second $U$-spin relation in (\ref{U-spin-rel}).

This strategy is very promising from an experimental point of view for LHCb, 
where an accuracy for $\gamma$ of a few degrees can be achieved 
\cite{LHCb-analyses}. As far as possible $U$-spin-breaking 
corrections to $d'=d$ are concerned, they enter the determination of $\gamma$ 
through a relative shift of the $\gamma$--$d$ and $\gamma$--$d'$ contours; 
their impact on the extracted value of $\gamma$ therefore depends on the form 
of these curves, which is fixed through the measured observables. In the examples discussed in \cite{RF-BsKK} and Fig.~\ref{fig:Bs-Bd-contours}, the extracted value 
of $\gamma$ would be very stable with respect to such effects. It should also be noted
that the $U$-spin relations in (\ref{U-spin-rel}) are particularly robust since they 
involve only ratios of hadronic amplitudes, where all $SU(3)$-breaking decay constants
and form factors cancel in factorization and also chirally enhanced terms
would not lead to  $U$-spin-breaking corrections \cite{RF-BsKK}. 

As a by-product of the $B\to\pi\pi,\pi K$ strategy developed in \cite{BFRS}, the
observables of the $B^0_s\to K^+K^-$ decay can be predicted in the SM. The most
recent data yield the following numbers \cite{FRS-07}:
\begin{eqnarray}%\label{}
{\cal A}_{\rm CP}^{\rm dir}(B_s\to K^+K^-)|_{\rm SM}&=&
0.093\pm0.015\\
{\cal A}_{\rm CP}^{\rm mix}(B_s\to K^+K^-)|_{\rm SM}&=&-0.234_{-0.014}^{+0.017}.
\end{eqnarray}
In the case of the CP-averaged branching ratio, an $SU(3)$-breaking form-factor
ratio enters the prediction, thereby increasing the uncertainties. Using the
result of a QCD sum-rule calculation \cite{Khod} yields the prediction \cite{FRS-07}
\begin{equation}\label{BsKK-BR}
\mbox{BR}(B_s\to K^+K^-)=(28_{-5}^{+7})\times 10^{-6}.
\end{equation}
The $B^0_s\to K^+K^-$ mode was recently observed by CDF \cite{CDF-BsK+K-}; 
the most recent experimental update for the CP-averaged branching ratio reads 
as follows \cite{CDF-punzi}: 
\begin{equation}\label{BsKK-exp}
\mbox{BR}(B_s\to K^+K^-)=(24.4\pm1.4\pm4.6)\times10^{-6}.
\end{equation}
Within the uncertainties, (\ref{BsKK-BR}) is in nice agreement with (\ref{BsKK-exp}),
which is another support of the working hypotheses underlying the $B\to\pi K$ analysis
discussed in Section~\ref{ssec:BpiK}.

\boldmath
\subsubsection{The Rare Decays $B_{s,d}\to\mu^+\mu^-$}
\unboldmath
In the SM, these decays originate from $Z$ penguins and box diagrams, and
the corresponding low-energy effective Hamiltonian takes the following form
\cite{BBL}:
\begin{equation}\label{Heff-Bmumu}
{\cal H}_{\rm eff}=-\frac{G_{\rm F}}{\sqrt{2}}\left[
\frac{\alpha}{2\pi\sin^2\Theta_{\rm W}}\right]
V_{tb}^\ast V_{tq} \eta_Y Y_0(x_t)(\bar b q)_{\rm V-A}(\bar\mu\mu)_{\rm V-A} 
\,+\, {\rm h.c.},
\end{equation}
where $\alpha$ denotes the QED coupling and $\Theta_{\rm W}$ is the
Weinberg angle. The short-distance physics is described by 
$Y(x_t)\equiv\eta_Y Y_0(x_t)$, where $\eta_Y=1.012$ is a perturbative 
QCD correction \cite{Bmumu}, and the Inami--Lim function
$Y_0(x_t)$ describes the top-quark mass dependence. We observe that
only the matrix element $\langle 0| (\bar b q)_{\rm V-A}|B^0_q\rangle$ 
is required. Since here the vector-current piece vanishes, as
the $B^0_q$ is a pseudoscalar meson, this matrix element is simply
given by the decay constant $f_{B_q}$. 
Consequently, we arrive at a very favourable 
situation with respect to the hadronic matrix elements. Since, moreover, 
NLO QCD corrections were calculated, and long-distance contributions are 
expected to play a negligible r\^ole \cite{Bmumu}, the $B^0_q\to\mu^+\mu^-$ 
modes belong to the cleanest rare $B$ decays. 

Using also the data for the mass differences $\Delta M_q$ to reduce the hadronic 
uncertainties,\footnote{This input allows us to replace the decay constants $f_{B_q}$
through the bag parameters $\hat B_{B_q}$.} the following SM predictions were 
obtained in \cite{BBGT}:
\begin{eqnarray}
\mbox{BR}(B_s\to\mu^+\mu^-) &=& (3.35\pm0.32)\times 10^{-9}\\
\mbox{BR}(B_d\to\mu^+\mu^-) &=& (1.03\pm0.09)\times 10^{-10}.
\end{eqnarray}
The upper bounds (95\% C.L.) from the CDF collaboration read as follows
\cite{CDF-Bmumu}:
\begin{equation}\label{Bmumu-exp-CDF}
\mbox{BR}(B_s\to\mu^+\mu^-)<1.0\times10^{-7}, \quad
\mbox{BR}(B_d\to\mu^+\mu^-)<3.0 \times10^{-8},
\end{equation}
while the D0 collaboration finds the following 90\% C.L. (95\% C.L.) upper limit 
\cite{D0-Bmumu}:
\begin{equation}\label{Bmumu-exp-D0}
\mbox{BR}(B_s\to\mu^+\mu^-)<1.9~(2.3) \times 10^{-7}.
\end{equation}
Consequently, there is still a long way to go within the SM. However, in this case, 
LHCb expects a $3\,\sigma$ observation for $B_s\to\mu^+\mu^-$ with already
nominal one year data ($2\,\mbox{fb}^{-1}$) \cite{nakada}. This decay is also very 
interesting for ATLAS and CMS, where detailed background studies are currently
in progress \cite{smsp}. Things could actually be much more exciting, as NP 
effects may significantly enhance $\mbox{BR}(B_s\to\mu^+\mu^-)$. For instance,
in SUSY, this enhancement may be dramatic as $\mbox{BR}\sim(\tan\beta)^6$,
where $\beta$ is here the ratio of the two Higgs vacuum expectation values and
not the UT angle $\beta$ (for recent analyses, see, e.g., \cite{Bmumu-recent}), and
in scenarios with a modified EW penguin sector a sizeable enhancement is
possible (see, e.g., \cite{BFRS}).

\boldmath
\subsubsection{The Rare Decay $B^0_d\to K^{*0}\mu^+\mu^-$}
\unboldmath
The key observable for NP searches provided by this decay is the following
forward--backward asymmetry:
\begin{equation}%\label{}
A_{\rm FB}(\hat s)=\frac{1}{{\rm d}\Gamma/{\rm d}\hat s}
\left[\int_0^{+1} {\rm d}(\cos\theta)\frac{{\rm d}^2\Gamma}{{\rm d} \hat s \, 
{\rm d}(\cos\theta)} - \int_{-1}^0{\rm d}(\cos\theta)
\frac{{\rm d}^2\Gamma}{{\rm d} \hat s \, {\rm d}(\cos\theta)}\right].
\end{equation}
Here $\theta$ is the angle between the $B^0_d$ momentum and that of the 
$\mu^+$ in the dilepton centre-of-mass system, and $\hat s \equiv s/M_B^2$ with 
$s=(p_{\mu^+}+p_{\mu^-})^2$. A particularly interesting kinematical point is
characterized by  
\begin{equation}%\label{}
A_{\rm FB}(\hat s_0)|_{\rm SM}=0,
\end{equation}
as $\hat s_0$ is quite robust with respect to hadronic uncertainties 
(see, e.g., \cite{BKastll}). In SUSY extensions of the SM, $A_{\rm FB}(\hat s)$ 
could take opposite sign or take a dependence on $\hat s$ without a zero point 
\cite{ABHH}. The current $B$-factory data for the inclusive $b\to s\ell^+\ell^-$ 
branching ratios and the integrated forward--backward asymmetries are in 
accordance with the SM, but suffer still from large uncertainties. This situation will 
improve dramatically at the LHC. Here LHCb will collect about $4400$ decays/year, 
yielding $\Delta \hat s_0= 0.06$ after one year, and ATLAS expects to collect about
1000 $B^0\to K^{*0}\mu^+\mu^-$ decays per year \cite{schopper}. Moreover, also 
other $b\to s\mu^+\mu^-$ modes are currently under study, such as 
$\Lambda_b\to \Lambda\mu^+\mu^-$ and $B^0_s\to\phi\mu^+\mu^-$.

\section{Conclusions and Outlook}
We have seen tremendous progress in $B$ physics during the recent years,
which was made possible through a fruitful interplay between theory and
experiment. Altogether, the $e^+e^-$ $B$ factories have already produced 
${\cal O}(10^9)$ $B\bar B$ pairs, and the Tevatron has recently succeeded in 
observing $B^0_s$--$\bar B^0_s$ mixing. The data agree globally with the KM 
mechanism of CP violation in an impressive manner, but we have also hints for 
discrepancies, which could be first signals of NP. Unfortunately, definite conclusions 
cannot yet be drawn as the uncertainties are still too large. 

Exciting new perspectives for $B$ physics and the exploration of CP violation 
will emerge through the start of the LHC in the autumn of 2007, with its
dedicated $B$-decay experiment LHCb. Thanks to the large statistics that
can be collected there and the full exploitation of the physics potential of the
$B_s$-meson system, we will be able to enter a new territory in the exploration
of CP violation at the LHC. The golden channel to search for CP-violating NP 
contributions to $B^0_s$--$\bar B^0_s$ mixing is $B^0_s\to J/\psi \phi$, where the
recent measurement of $\Delta M_s$ still leaves ample space for such effects both in 
terms of the general NP parameters and in specific extensions of the SM. In contrast
to the theoretical interpretation of $\Delta M_s$, the corresponding CP asymmetries
have not to rely on non-perturbative lattice QCD calculations. The two major lines of
the broad research programme of LHCb are precision measurements of $\gamma$,
which is a key ingredient for NP searches, and powerful analyses of various
rare $B$ decays, offering also sensitive probes for physics beyond the SM. The 
implementation of this programme will lead to much more stringent consistency
checks of the KM mechanism, where also measurements of the rare kaon decays
$K^+\to\pi^+\nu\bar\nu$ and $K_{\rm L}\to\pi^0\nu\bar\nu$ would be very welcome.

These studies of CP violation and flavour physics play also an outstanding r\^ole 
in the context with the major targets of the physics programme of the LHC. Here the 
main goal of the ATLAS and CMS experiments is to explore electroweak 
symmetry breaking, in particular the question of whether this is actually caused by 
the Higgs mechanism, to produce and observe new particles, and then to go
back to the deep questions of particle physics, such as the origin of dark matter
and the baryon asymmetry of the Universe. It is obvious that there should be a 
very fruitful interplay between these ``direct" studies of NP and the ``indirect" 
information provided by flavour physics, including the $B$-meson system, 
but also $K$, $D$ and top physics as well as the flavour physics in the lepton 
sector.\footnote{This synergy is the topic of a CERN Workshop: 
http://flavlhc.web.cern.ch/flavlhc/.} I have no doubts that the next years will be
extremely exciting!

\bigskip
I am grateful to the workshop organizers for the invitation to this interesting meeting 
at such a nice location, and would also like to thank my co-authors for the enjoyable 
collaborations on topics addressed in this talk.


\begin{thebibliography}{99}

%%
%%  bibliographic items can be constructed using the LaTeX format in SPIRES:
%%    see    http://www.slac.stanford.edu/spires/hep/latex.html
%%  SPIRES will also supply the CITATION line information; please include it.
%%

\bibitem{KM}M.~Kobayashi and T.~Maskawa,
  %``CP Violation In The Renormalizable Theory Of Weak Interaction,''
  Prog.\ Theor.\ Phys.\  {\bf 49}, 652 (1973).
  %%CITATION = PTPKA,49,652;%%

\bibitem{BPY}W.~Buchm\"uller, R.~D.~Peccei and T.~Yanagida,
  %``Leptogenesis as the origin of matter,''
  Ann.\ Rev.\ Nucl.\ Part.\ Sci.\  {\bf 55}, 311 (2005).
 % [arXiv:hep-ph/0502169].
  %%CITATION = ARNUA,55,311;%%

\bibitem{buja}A.~J.~Buras and M.~Jamin,
  %``epsilon'/epsilon at the NLO: 10 years later,''
  { JHEP} {\bf 0401}, 048 (2004).
 % [arXiv:hep-ph/0306217].
  %%CITATION = HEP-PH 0306217;%%

\bibitem{CPV}J.~H.~Christenson {\it et al.}, 
%``Evidence For The 2 Pi Decay Of The K(2)0 Meson,''
{Phys.\ Rev.\ Lett.}~{\bf 13}, 138 (1964).
%%CITATION = PRLTA,13,138;%%

\bibitem{BSU}A.~J.~Buras, F.~Schwab and S.~Uhlig,
  %``Waiting for precise measurements of K+ --> pi+ nu anti-nu and K(L) -->  pi0
  %nu anti-nu,''
  hep-ph/0405132.
  %%CITATION = HEP-PH/0405132;%%
  
\bibitem{Ruggiero}G. Ruggiero, talk at this workshop.
  
\bibitem{Iijima}T. Iijima, talk at this workshop.

\bibitem{Belle-U5S}K.~Abe {\it et al.}  [Belle Collaboration],
  %``Measurements of exclusive B/s0 decays at the Upsilon(5S),''
  hep-ex/0610003.
  %%CITATION = HEP-EX/0610003;%%
  
\bibitem{wolf-blo}L.~Wolfenstein,
  %``Parametrization Of The Kobayashi-Maskawa Matrix,''
  Phys.\ Rev.\ Lett.\  {\bf 51}, 1945 (1983);
  %%CITATION = PRLTA,51,1945;%%
  A.~J.~Buras, M.~E.~Lautenbacher and G.~Ostermaier,
  %``Waiting For The Top Quark Mass, K+ $\to$ Pi+ Neutrino Anti-Neutrino, B(S)0
  %- Anti-B(S)0 Mixing And CP Asymmetries In B Decays,''
  Phys.\ Rev.\  D {\bf 50}, 3433 (1994).
 % [arXiv:hep-ph/9403384].
  %%CITATION = PHRVA,D50,3433;%%

\bibitem{BBL}G.~Buchalla, A.~J.~Buras and M.~E.~Lautenbacher,
%``Weak Decays Beyond Leading Logarithms,''
{ Rev.\ Mod.\ Phys.}~{\bf 68}, 1125 (1996).
%%CITATION = HEP-PH 9512380;%%

\bibitem{beneke}M.~Beneke, talk at this workshop 
%``Hadronic B decays,''
  [hep-ph/0612353].
  %%CITATION = HEP-PH/0612353;%%

\bibitem{gw}M. Gronau and D. Wyler,
%``On determining a weak phase from CP asymmetries in charged B decays,''
{ Phys.\ Lett.}\ B {\bf 265}, 172 (1991).
%%CITATION = PHLTA,B265,172;%%

\bibitem{ADS}D. Atwood, I. Dunietz and A. Soni,
 %``Enhanced CP violation with B $\to$ K D0 (anti-D0) modes and extraction  of
%the CKM angle gamma,''
{ Phys.\ Rev.\ Lett.}~{\bf 78}, 3257 (1997);
%[arXiv:hep-ph/9612433].
%%CITATION = HEP-PH 9612433;%%
 %``Improved methods for observing CP violation in B+- $\to$ K D and  measuring
%the CKM phase gamma,''
{ Phys.\ Rev.}\ D  {\bf 63}, 036005 (2001).
%[arXiv:hep-ph/0008090].
%%CITATION = HEP-PH 0008090;%%

\bibitem{fw}R. Fleischer and D. Wyler,
%``Exploring CP violation with B/c decays,''
{ Phys.\ Rev.}\ D {\bf 62}, 057503 (2000).
%[arXiv:hep-ph/0004010].
%%CITATION = HEP-PH 0004010;%%

\bibitem{GHLR}M.~Gronau, J.~L.~Rosner and D.~London,
  %``Weak coupling phase from decays of charged B mesons to pi K and pi pi,''
  { Phys.\ Rev.\ Lett.}~{\bf 73}, 21 (1994);
  %[arXiv:hep-ph/9404282].
  %%CITATION = HEP-PH 9404282;%%
 M.~Gronau, O.~F.~Hernandez, D.~London and J.~L.~Rosner,
  %``Decays of B mesons to two light pseudoscalars,''
  { Phys.\ Rev.}\ D {\bf 50}, 4529 (1994).
  %[arXiv:hep-ph/9404283].
  %%CITATION = HEP-PH 9404283;%%
  
\bibitem{bisa}A.~B.~Carter and A.~I.~Sanda,
{ Phys.\ Rev.\ Lett.}~{\bf 45}, 952 (1980);
%%CITATION = PRLTA,45,952;%%
{ Phys.\ Rev.}\ D {\bf 23}, 1567 (1981);
%%CITATION = PHRVA,D23,1567;%%
I.~I.~Bigi and A.~I.~Sanda,
{ Nucl.\ Phys.}\ B {\bf 193}, 85 (1981).
%%CITATION = NUPHA,B193,85;%%

\bibitem{RF-lect}R.~Fleischer,
  %``Flavour physics and CP violation,''
  lectures given at the 2005 European School of High-Energy Physics, Kitzb\"uhel, 
  Austria, 21 August -- 3 September 2005 [hep-ph/0608010].
  %%CITATION = HEP-PH/0608010;%%
  
\bibitem{weiler}A. Weiler, talk at this workshop.

\bibitem{stocchi}A. Stocchi, talk at this workshop.

\bibitem{CKMfitter}J.~Charles {\it et al.}~[CKMfitter Group], 
{Eur.\ Phys.\ J.}~C {\bf 41}, 1 (2005); for the most recent updates, see
http://ckmfitter.in2p3.fr/.

\bibitem{UTfit}M.~Bona {\it et al.}~[UTfit Collaboration],
  %``The 2004 UTfit collaboration report on the status of the unitarity triangle
  %in the standard model,''
  {JHEP} {\bf 0507}, 028 (2005); for the most recent updates, see
http://utfit.roma1.infn.it/.
  %[arXiv:hep-ph/0501199].
  %%CITATION = HEP-PH 0501199;%%
  
\bibitem{RF-EWP-rev}R. Fleischer,
  %``CP violation and the role of electroweak penguins in non-leptonic B
  %decays,''
  { Int.\ J.\ Mod.\ Phys.}\ A {\bf 12}, 2459 (1997).
  %[arXiv:hep-ph/9612446].
  %%CITATION = HEP-PH 9612446;%%

\bibitem{growo}Y.~Grossman and M.~P.~Worah,
  %``CP asymmetries in B decays with new physics in decay amplitudes,''
  { Phys.\ Lett.}\ B {\bf 395}, 241 (1997).
  %[arXiv:hep-ph/9612269].
  %%CITATION = HEP-PH 9612269;%%

\bibitem{FM-BphiK}R. Fleischer and T. Mannel,
 %``Exploring new physics in the B $\to$ Phi K system,''
 { Phys.\ Lett.}\ B {\bf 511}, 240 (2001).
 %[arXiv:hep-ph/0103121].
 %%CITATION = HEP-PH 0103121;%%

\bibitem{BphiK-NP}M. Ciuchini, E. Franco, A. Masiero and L. Silvestrini,
  %``Worries and hopes for SUSY in CKM physics: The b $\to$ s example,''
  { eConf} {\bf C0304052}, WG307 (2003); 
%[arXiv:hep-ph/0308013].
  %%CITATION = HEP-PH 0308013;%%
  V. Barger, C.~W. ~Chiang, P. Langacker and H.~S.~Lee,
  %``Solution to the B $\to$ pi K puzzle in a flavor-changing Z' model,''
  { Phys.\ Lett.}\ B {\bf 598}, 218 (2004).
  %[arXiv:hep-ph/0406126].
  %%CITATION = HEP-PH 0406126;%%
  
\bibitem{DuFe}G. Dubois--Felsmann, talk at this workshop.

\bibitem{HFAG}Heavy Flavour Averaging Group [E. Barberio {\it et al.}], 
hep-ex/0603003; for online updates, see  http://www.slac.stanford.edu/xorg/hfag. 
  %%CITATION = HEP-EX 0603003;%%

\bibitem{BpiK-papers}T.~Yoshikawa,
  %``A possibility of large electro-weak penguin contribution in B --> K pi
  %modes,''
  Phys.\ Rev.\ D {\bf 68}, 054023 (2003);
  %[arXiv:hep-ph/0306147].
  %%CITATION = HEP-PH 0306147;%%
  M.~Gronau and J.~L.~Rosner,
  %``Rates and asymmetries in B --> K pi decays,''
  Phys.\ Lett.\ B {\bf 572}, 43 (2003);
  %[arXiv:hep-ph/0307095].
  %%CITATION = HEP-PH 0307095;%%
  M.~Beneke and M.~Neubert,
  %``QCD factorization for B --> P P and B --> P V decays,''
  Nucl.\ Phys.\ B {\bf 675}, 333 (2003);
 % [arXiv:hep-ph/0308039].
  %%CITATION = HEP-PH 0308039;%%
  V.~Barger, C.~W.~Chiang, P.~Langacker and H.~S.~Lee,
  %``Solution to the B --> pi K puzzle in a flavor-changing Z' model,''
  Phys.\ Lett.\ B {\bf 598}, 218 (2004);
 % [arXiv:hep-ph/0406126].
  %%CITATION = HEP-PH 0406126;%%
   Y.~L.~Wu and Y.~F.~Zhou,
  %``Charmless decays B --> pi pi, pi K and K K in broken SU(3) symmetry,''
  Phys.\ Rev.\ D {\bf 72}, 034037 (2005).
  %[arXiv:hep-ph/0503077].
  %%CITATION = HEP-PH 0503077;%%

\bibitem{FRS-07}R.~Fleischer, S.~Recksiegel and F.~Schwab,
  %``On puzzles and non-puzzles in B --> pi pi, pi K decays,''
  CERN-PH-TH/2007-044, to appear in Eur.\ Phys.\ J. C [hep-ph/0702275].
  %%CITATION = HEP-PH/0702275;%%

\bibitem{BFRS}A.~J.~Buras, R.~Fleischer, S.~Recksiegel and F.~Schwab,
  %``B --> pi pi, new physics in B --> pi K and implications for rare K and  B
  %decays,''
  Phys.\ Rev.\ Lett.\  {\bf 92}, 101804 (2004);
  %[arXiv:hep-ph/0312259].
  %%CITATION = HEP-PH 0312259;%%
  %``Anatomy of prominent B and K decays and signatures of CP-violating new
  %physics in the electroweak penguin sector,''
  Nucl.\ Phys.\ B {\bf 697}, 133 (2004);
  %[arXiv:hep-ph/0402112].
  %%CITATION = HEP-PH 0402112;%%
%``New aspects of B --> pi pi, pi K and their implications for rare decays,''
  Eur.\ Phys.\ J.\ C {\bf 45}, 701 (2006).
  %[arXiv:hep-ph/0512032].
  
\bibitem{BF98}A.~J.~Buras and R.~Fleischer,
  %``A general analysis of gamma determinations from B --> pi K decays,''
  Eur.\ Phys.\ J.\  C {\bf 11}, 93 (1999).
 % [arXiv:hep-ph/9810260].
  %%CITATION = EPHJA,C11,93;%%
  
\bibitem{NR}M.~Neubert and J.~L.~Rosner,
  %``Determination of the weak phase gamma from rate measurements in  B+- --> pi
  %K, pi pi decays,''
  Phys.\ Rev.\ Lett.\  {\bf 81}, 5076 (1998).
 % [arXiv:hep-ph/9809311].
  %%CITATION = HEP-PH 9809311;%%

\bibitem{EWP-NP}R.~Fleischer and T.~Mannel,
  %``New physics in penguin dominated B --> pi K decays,''
  hep-ph/9706261;
  %%CITATION = HEP-PH 9706261;%%
Y.~Grossman, M.~Neubert and A.~L.~Kagan,
  %``Trojan penguins and isospin violation in hadronic B decays,''
  JHEP {\bf 9910}, 029 (1999).
 % [arXiv:hep-ph/9909297].
  %%CITATION = HEP-PH 9909297;%%

\bibitem{BF00}A.~J.~Buras and R.~Fleischer,
  %``Constraints on the CKM angle gamma and strong phases from B --> pi K
  %decays,''
  Eur.\ Phys.\ J.\  C {\bf 16}, 97 (2000).
  %[arXiv:hep-ph/0003323].
  %%CITATION = EPHJA,C16,97;%%

\bibitem{GR-06}M.~Gronau and J.~L.~Rosner,
  %``Rate and CP-asymmetry sum rules in B --> K pi,''
  Phys.\ Rev.\  D {\bf 74}, 057503 (2006);
 % [arXiv:hep-ph/0608040].
  %%CITATION = PHRVA,D74,057503;%%
  M.~Gronau and J.~L.~Rosner,
  %``Sum rule for rate and CP asymmetry in B+ --> K+ pi0,''
  Phys.\ Lett.\  B {\bf 644}, 237 (2007).
  %[arXiv:hep-ph/0610227].
  %%CITATION = PHLTA,B644,237;%%

\bibitem{BF-06}P.~Ball and R.~Fleischer,
  %``Probing new physics through B mixing: Status, benchmarks and prospects,''
  Eur.\ Phys.\ J.\  C {\bf 48}, 413 (2006).
  %[arXiv:hep-ph/0604249].
  %%CITATION = EPHJA,C48,413;%%
  
\bibitem{lubicz}V. Lubicz, talk at this workshop
%``Lattice QCD, flavor physics and the unitarity triangle analysis,''
  [hep-ph/0702204].
  %%CITATION = HEP-PH/0702204;%%

\bibitem{JLQCD}S.~Aoki {\it et al.}\  [JLQCD collaboration],
  %``B0 anti-B0 mixing in unquenched lattice QCD,''
{ Phys.\ Rev.\ Lett.} {\bf 91}, 212001 (2003).
%  [arXiv:hep-ph/0307039].
  %%CITATION = HEP-PH 0307039;%%

\bibitem{HPQCD}A.~Gray {\it et al.}\  [HPQCD collaboration],
  %``The B meson decay constant from unquenched lattice QCD,''
{  Phys.\ Rev.\ Lett.} {\bf 95}, 212001 (2005).
%  [arXiv:hep-lat/0507015].
  %%CITATION = HEP-LAT 0507015;%%

\bibitem{Okamoto}M.~Okamoto,
  %``Full determination of the CKM matrix using recent results from lattice
  %QCD,''
 { PoS {\bf LAT2005}}, 013 (2005).
%  [arXiv:hep-lat/0510113].
  %%CITATION = HEP-LAT 0510113;%%
  
\bibitem{Yamamoto}H. Yamamoto, talk at this workshop.

\bibitem{Buchmuller}O. Buchm\"uller, talk at this workshop. 

\bibitem{Mannel}T. Mannel, talk at this workshop.

\bibitem{lenz}A. Lenz, talk at this workshop 
  %``Lifetimes and oscillations of heavy mesons,''
  [hep-ph/0612176].
  %%CITATION = HEP-PH/0612176;%%

\bibitem{DDF}A.~S.~Dighe, I.~Dunietz and R.~Fleischer,
%``Extracting CKM phases and B/s anti-B/s mixing parameters from  
%angular distributions of non-leptonic B decays,''
{ Eur.\ Phys.\ J.}\ C {\bf 6}, 647 (1999).
%%CITATION = HEP-PH 9804253;%%

\bibitem{CDF-DG}D. Acosta {\it et al.}\  [CDF Collaboration],
  %``Analysis of decay-time dependence of angular distributions in B/s0 $\to$
  %J/psi Phi and B/d0 $\to$ J/psi K*0 decays and measurement of the lifetime
  %difference between B/s mass eigenstates,''
  { Phys.\ Rev.\ Lett.}~{\bf 94}, 101803 (2005).
  %[arXiv:hep-ex/0412057].
  %%CITATION = HEP-EX 0412057;%%

\bibitem{D0-DG}V.~M.~Abazov {\it et al.}\  [D0 Collaboration],
  %``Measurement of the lifetime difference in the B/s0 system,''
  { Phys.\ Rev.\ Lett.}~{\bf 95}, 171801 (2005).
%  [arXiv:hep-ex/0507084].
  %%CITATION = HEP-EX 0507084;%%
  
\bibitem{schopper}A.~Schopper,
  %``Flavor physics and CP violation at LHC,''
Proceedings of FPCP 2006, Vancouver, British Columbia, Canada, 9--12 April 
2006, pp 042  [hep-ex/0605113].
  %%CITATION = ECONF,C060409,042;%%
  
\bibitem{nakada}T.~Nakada, talk at CKM 2006, Nagoya, Japan, 
12--16 December 2006. 

\bibitem{dun}I. Dunietz,
 %``B(s) - anti-B(s) mixing, CP violation and extraction of CKM phases from
%untagged B(s) data samples,''
{ Phys.\ Rev.}\ D {\bf 52}, 3048 (1995).
%[arXiv:hep-ph/9501287].
%%CITATION = HEP-PH 9501287;%%

\bibitem{FD-CP}R. Fleischer and I. Dunietz,
 %``CP violation and CKM phases from angular distributions for $B_s$ 
%decays into
%admixtures of CP eigenstates,''
{ Phys.\ Rev.}\ D {\bf 55}, 259 (1997).
%[arXiv:hep-ph/9605220].
%%CITATION = HEP-PH 9605220;%%

\bibitem{FD-NCP}R. Fleischer and I. Dunietz,
 %``CP violation and the CKM angle $\gamma$ from angular distributions of
%untagged $B_s$ decays governed by $\bar b\to\bar c u\bar s$,''
{ Phys.\ Lett.}\ B {\bf 387}, 361 (1996).
%[arXiv:hep-ph/9605221].
%%CITATION = HEP-PH 9605221;%%

\bibitem{DFN}I. Dunietz, R. Fleischer and U. Nierste,
  %``In pursuit of new physics with B/s decays,''
  { Phys.\ Rev.}\ D {\bf 63}, 114015 (2001).
  %[arXiv:hep-ph/0012219].
  %%CITATION = HEP-PH 0012219;%%

\bibitem{NiSi}Y.~Nir and D.~J.~Silverman,
  %``EXPLORING NEW PHYSICS WITH CP ASYMMETRIES IN B0 DECAYS,''
  Nucl.\ Phys.\  B {\bf 345}, 301 (1990).
  %%CITATION = NUPHA,B345,301;%%

\bibitem{BMPR}G.~C.~Branco, T.~Morozumi, P.~A.~Parada and M.~N.~Rebelo,
  %``CP asymmetries in B0 decays in the presence of flavor changing neutral
  %currents,''
  Phys.\ Rev.\  D {\bf 48}, 1167 (1993).
  %%CITATION = PHRVA,D48,1167;%%

\bibitem{Bosc-WG}$B$ Oscillations Working Group: 
http://lepbosc.web.cern.ch/LEPBOSC/.

\bibitem{menzemer}S. Menzemer, talk at this workshop.

\bibitem{D0} V.~M.~Abazov {\it et al.}\  [D0 Collaboration],
  %``First direct two-sided bound on the B/s0 oscillation frequency,''
   Phys.\ Rev.\ Lett.\  {\bf 97}, 021802 (2006).
  %[arXiv:hep-ex/0603029].
  %%CITATION = PRLTA,97,021802;%%
  
\bibitem{CDF}A.~Abulencia {\it et al.}  [CDF Collaboration],
  %``Observation of B/s0 anti-B/s0 oscillations,''
  Phys.\ Rev.\ Lett.\  {\bf 97}, 242003 (2006).
  %[arXiv:hep-ex/0609040].
  %%CITATION = PRLTA,97,242003;%%
  
\bibitem{DMs-papers}M.~Carena {\it et al.}, %A.~Menon, R.~Noriega-Papaqui, A.~Szynkman and C.E.M.~Wagner,
  %``Constraints on B and Higgs physics in minimal low energy supersymmetric
  %models,''
   Phys.\ Rev.\  D {\bf 74}, 015009 (2006);
 % [arXiv:hep-ph/0603106].
  %%CITATION = PHRVA,D74,015009;%%
  M.~Ciuchini and L.~Silvestrini,
  %``Upper bounds on SUSY contributions to b $\to$ s transitions from B/s -
  %anti-B/s mixing,''
  Phys.\ Rev.\ Lett.\  {\bf 97}, 021803 (2006);
 % [arXiv:hep-ph/0603114].
  %%CITATION = PRLTA,97,021803;%%
 M.~Endo and S.~Mishima,
  %``Constraint on right-handed squark mixings from B/s - anti-B/s mass
  %difference,''
  Phys.\ Lett.\  B {\bf 640}, 205 (2006);
 % [arXiv:hep-ph/0603251].
  %%CITATION = PHLTA,B640,205;%%
  Z.~Ligeti, M.~Papucci and G.~Perez,
  %``Implications of the measurement of the B^0_s-\bar B^0_s mass difference,''
  Phys.\ Rev.\ Lett.\  {\bf 97}, 101801 (2006);
  %[arXiv:hep-ph/0604112].
  %%CITATION = PRLTA,97,101801;%%
  J.~Foster, K.I.~Okumura and L.~Roszkowski,
  %``New constraints on SUSY flavour mixing in light of recent measurements at
  %the Tevatron,''
   Phys.\ Lett.\  B {\bf 641}, 452 (2006);
  %[arXiv:hep-ph/0604121].
  %%CITATION = PHLTA,B641,452;%%.
 Y.~Grossman, Y.~Nir and G.~Raz,
  %``Constraining the phase of B/s - anti-B/s mixing,''
  Phys.\ Rev.\ Lett.\  {\bf 97}, 151801 (2006);
 % [arXiv:hep-ph/0605028].
  %%CITATION = PRLTA,97,151801;%%
S.~Baek, J.~H.~Jeon and C.~S.~Kim,
  %``B/s0 - anti-B/s0 mixing in leptophobic Z' model,''
  Phys.\ Lett.\  B {\bf 641}, 183 (2006);
 % [arXiv:hep-ph/0607113].
  %%CITATION = PHLTA,B641,183;%%
  M.~Blanke and A.~J.~Buras,
  %``Lower bounds on Delta(M/s,d) from constrained minimal flavour violation,''
  hep-ph/0610037.
  %%CITATION = HEP-PH/0610037;%%
  
\bibitem{BBGT}M.~Blanke, A.~J.~Buras, D.~Guadagnoli and C.~Tarantino,
  %``Minimal flavour violation waiting for precise measurements of Delta(M(s)),
  %|V(ub)|, gamma and B/s,d0 --> mu+ mu-,''
  JHEP {\bf 0610}, 003 (2006).
 % [arXiv:hep-ph/0604057].
  %%CITATION = JHEPA,0610,003;%%

\bibitem{HPQCD-DMs}E.~Dalgic {\it et al.}\  [HPQCD collaboration],
  %``B/s0 - anti-B/s0 mixing parameters from unquenched lattice QCD,''
  hep-lat/0610104.
  %%CITATION = HEP-LAT/0610104;%%

\bibitem{DDLR}A.~S.~Dighe, I.~Dunietz, H.~J.~Lipkin and J.~L.~Rosner,
  %``Angular distributions and lifetime differences in $B_s \to J/\psi \phi$
  %decays,''
  Phys.\ Lett.\  B {\bf 369}, 144 (1996).
  %[arXiv:hep-ph/9511363].
  %%CITATION = PHLTA,B369,144;%%

\bibitem{D0-phis}V.~M.~Abazov {\it et al.}  [D0 Collaboration],
  %``Lifetime difference and CP-violating phase in the B/s0 system,''
  hep-ex/0701012.
  %%CITATION = HEP-EX/0701012;%%

\bibitem{smsp}Maria Smizanska  and Thomas Speer, private communications.

\bibitem{LHCb-up}F.~Muheim,
  %``LHCb Upgrade Plans,''
  hep-ex/0703006.
  %%CITATION = HEP-EX/0703006;%%

\bibitem{RF-ang}R.~Fleischer,
  %``Extracting CKM phases from angular distributions of B/d,s decays into
  %admixtures of CP eigenstates,''
  Phys.\ Rev.\  D {\bf 60}, 073008 (1999).
  %[arXiv:hep-ph/9903540].
  %%CITATION = PHRVA,D60,073008;%%

\bibitem{LH}M.~Blanke {\it et al.}, 
%A.~J.~Buras, A.~Poschenrieder, S.~Recksiegel, C.~Tarantino, S.~Uhlig and A.~Weiler,
  %``Rare and CP-violating K and B decays in the littlest Higgs model with
  %T-parity,''
  JHEP {\bf 0701}, 066 (2007);
  %[arXiv:hep-ph/0610298].
  %%CITATION = JHEPA,0701,066;%%
  C.~Tarantino, talk at this workshop
  %``Flavour physics in the littlest Higgs model with T-parity,''
  [hep-ph/0702152].
  %%CITATION = HEP-PH/0702152;%%

\bibitem{BsDsK}R. Aleksan, I. Dunietz and B. Kayser,
%``Determining the CP violating phase gamma,''
{ Z.\ Phys.}\ C {\bf 54}, 653 (1992).
%%CITATION = ZEPYA,C54,653;%%

\bibitem{BdDpi}I.~Dunietz and R.~G.~Sachs,
%``Asymmetry Between Inclusive Charmed And Anticharmed Modes In B0, 
%Anti-B0 Decay As A Measure Of CP Violation,''
{ Phys.\ Rev.}\ D {\bf 37}, 3186 (1988) [E: D {\bf 39}, 3515 (1989)];
%%CITATION = PHRVA,D37,3186;%%
I.~Dunietz,
%``Clean CKM information from B/d(t) $\to$ D*-+ pi+-,''
{ Phys.\ Lett.}\ B {\bf 427}, 179 (1998);
%%CITATION = HEP-PH 9712401;%%
D.~A.~Suprun, C.~W.~Chiang and J.~L.~Rosner,
%``Extraction of a weak phase from B $\to$ D(*) pi,''
{ Phys.\ Rev.}\ D {\bf 65}, 054025 (2002).
%%CITATION = HEP-PH 0110159;%%

\bibitem{RF-gam-ca}R. Fleischer,
%``New strategies to obtain insights into CP violation through B/s $\to$
  %D/s+- K-+, D/s*+- K-+, ... and B/d $\to$ D+- pi-+, D*+- pi-+, ... decays,''
 {  Nucl.\ Phys.}\ B {\bf 671}, 459 (2003).
  %[arXiv:hep-ph/0304027].
  %%CITATION = HEP-PH 0304027;%%

\bibitem{WG5-rep}G. Wilkinson, in G.~Cavoto {\it et al.},
  %``Angles from B decays with charm,''
  Proceedings of CKM 2005 (WG5), San Diego, California, 15--18
   March 2005 [hep-ph/0603019], 
  %%CITATION = HEP-PH 0603019;%%
and private communication. 

\bibitem{RF-BsKK}R. Fleischer,
  %``New strategies to extract beta and gamma from B/d $\to$ pi+ pi- and  B/s
  %$\to$ K+ K-,''
  { Phys.\ Lett.}\ B {\bf 459}, 306 (1999).
  %[arXiv:hep-ph/9903456].
  %%CITATION = HEP-PH 9903456;%%

\bibitem{LHCb-analyses}G. Balbi {\it et al.}, CERN-LHCb/2003-123 and
124; R. Antunes Nobrega {\it et al.}\ [LHCb Collaboration], {\it Reoptimized
LHCb Detector, Design and Performance}, Technical Design Report 9, 
CERN/LHCC 2003-030;
J. Nardulli, talk at CKM 2006, Nagoya, Japan, 12--16 December 2006. 

\bibitem{Khod}A.~Khodjamirian, T.~Mannel and M.~Melcher,
  %``Kaon distribution amplitude from QCD sum rules,''
  Phys.\ Rev.\  D {\bf 70}, 094002 (2004).
 % [arXiv:hep-ph/0407226].
  %%CITATION = HEP-PH 0407226;%%

\bibitem{CDF-BsK+K-}A.~Abulencia {\it et al.}  [CDF Collaboration],
  %``Observation of B/s0 --> K+ K- and measurements of branching fractions  of
  %charmless two-body decays of B0 and B/s0 mesons in anti-p p  collisions at
  %s**(1/2) = 1.96-TeV,''
  { Phys.\ Rev.\ Lett.}~{\bf 97}, 211802 (2006). 
  %[arXiv:hep-ex/0607021].
  %%CITATION = PRLTA,97,211802;%%
  
\bibitem{CDF-punzi}G. Punzi, talk at CKM 2006, Nagoya, Japan, 
12--16 December 2006.

\bibitem{Bmumu}G.~Buchalla and A.~J.~Buras,
 %``The rare decays K $\to$ pi nu anti-nu, B $\to$ X nu anti-nu and  
%B $\to$ l+l-: An update,''
{ Nucl.\ Phys.}\ B {\bf 400}, 225 (1993) 
%[arXiv:hep-ph/9901288].
%%CITATION = HEP-PH 9901288;%%
and  B {\bf 548}, 309 (1999);
%%CITATION = NUPHA,B400,225;%%
M.~Misiak and J.~Urban, 
%``{QCD} corrections to FCNC decays mediated by Z-penguins and W-boxes,''
{ Phys.\ Lett.}\ B  {\bf 451}, 161 (1999).
%[arXiv:hep-ph/9901278].
%%CITATION = HEP-PH 9901278;%%

\bibitem{CDF-Bmumu}CDF Collaboration, CDF Public Note 8176 (2006)
[http://www-cdf.fnal.gov].

\bibitem{D0-Bmumu}D0 Collaboration, D0note 5009-CONF (2006) 
[http://www-d0.fnal.gov].

\bibitem{Bmumu-recent}J.~Foster, K.~Okumura and L.~Roszkowski,
  %``New constraints on SUSY flavour mixing in light of recent measurements at
  %the Tevatron,''
  Phys.\ Lett.\  B {\bf 641}, 452 (2006);
  %[arXiv:hep-ph/0604121].
  %%CITATION = PHLTA,B641,452;%%
  G.~Isidori and P.~Paradisi,
  %``Hints of large tan(beta) in flavour physics,''
  Phys.\ Lett.\  B {\bf 639}, 499 (2006).
 % [arXiv:hep-ph/0605012].
  %%CITATION = PHLTA,B639,499;%%
  
\bibitem{BKastll}G.~Burdman,
  %``Short distance coefficients and the vanishing of the lepton asymmetry  in B
  %--> V l+ l-,''
  Phys.\ Rev.\  D {\bf 57}, 4254 (1998);
  %[arXiv:hep-ph/9710550].
  %%CITATION = PHRVA,D57,4254;%%
  M.~Beneke, T.~Feldmann and D.~Seidel,
  %``Exclusive radiative and electroweak b --> d and b --> s penguin decays  at
  %NLO,''
  Eur.\ Phys.\ J.\  C {\bf 41}, 173 (2005);
  %[arXiv:hep-ph/0412400].
  %%CITATION = EPHJA,C41,173;%%
A.~Ali, G.~Kramer and G.~h.~Zhu,
  %``B --> K* l+ l- in soft-collinear effective theory,''
  Eur.\ Phys.\ J.\  C {\bf 47}, 625 (2006).
  %[arXiv:hep-ph/0601034].
  %%CITATION = EPHJA,C47,625;%%
  
\bibitem{ABHH}A.~Ali, P.~Ball, L.~T.~Handoko and G.~Hiller,
  %``A comparative study of the decays B --> (K,K*) l+ l- in standard  model and
  %supersymmetric theories,''
  Phys.\ Rev.\  D {\bf 61}, 074024 (2000).
 % [arXiv:hep-ph/9910221].
  %%CITATION = PHRVA,D61,074024;%%
  
  
\end{thebibliography}
\end{document}